\documentclass[a4paper,12pt, epsfig]{article}
\def\letter{0}\def\pr{0}
\usepackage{epsfig}
\usepackage{epstopdf}
\usepackage{textcomp}
\usepackage{graphicx}
\usepackage{ifthen}
\usepackage{ulem}

\usepackage{amsmath}
\usepackage[psamsfonts]{amssymb}
\usepackage{euscript}

\usepackage{latexsym}
\usepackage[arrow,matrix,curve]{xy}

\usepackage[hypertexnames=false]{hyperref}
\hypersetup{
    colorlinks=true,
    linkcolor=blue,
    filecolor=magenta,   
    citecolor=black,   
    urlcolor=cyan,
    }

\newskip\humongous \humongous=0pt plus 1000pt minus 1000pt

\newif\ifdtup

\def\,{\hspace{-.1cm}}
\def\hsp{,\hspace{.7cm}}

\def\fc#1#2 {\frac{n}{q}#1\frac{n}{q}#2}

\newcommand{\vac}{\ensuremath{|0\rangle}}

\renewcommand{\cos}{\textrm{cos}}
\renewcommand{\sin}{\textrm{sin}}

\renewcommand{\tanh}{\textrm{tanh}}
\newcommand{\sech}{\textrm{sech}}

\renewcommand{\cot}{\textrm{cot}}
\def\exp#1{\hbox{\rm exp}\left[#1\right]}

\renewcommand{\theequation}{\arabic{section}.\arabic{equation}}
\renewcommand{\(}{\begin{equation}}
\renewcommand{\)}{end{equation} \vspace{-.05in}\linebreak}

\newcounter{saveeqn}
\newcounter{savealpheqn}

\newcommand{\alpheqn}{\setcounter{saveeqn}{\value{equation}}%
  \stepcounter{saveeqn}\setcounter{equation}{0}%
  \renewcommand{\theequation}{\mbox{\arabic{section}.\arabic{saveeqn}
\alph{equation}}}
  \renewcommand{\)}{\end{equation}}}
\def\part#1{\frac{\partial}{\partial{#1}}}%
\def\group#1{\refstepcounter{equation}\setcounter{saveeqn}
 {\value{equation}}%
  \label{#1}\setcounter{equation}{0}%
\renewcommand{\theequation}{\mbox{\arabic{section}.\arabic{saveeqn}
\alph{equation}}}
  \renewcommand{\)}{\end{equation}}}
\newcommand{\reseteqn}{\setcounter{equation}{\value{saveeqn}}%
  \renewcommand{\theequation}{\arabic{section}.\arabic{equation}}%
  \renewcommand{\)}{\end{equation}}}

\newcommand{\aalpheqn}{\setcounter{saveeqn}{\value{equation}}%
  \stepcounter{saveeqn}\setcounter{equation}{0}%
  \renewcommand{\theequation}{\mbox{
        \Alph{subsection}.\arabic{saveeqn}\alph{equation}}}
   \renewcommand{\)}{\end{equation}}}
\newcommand{\areseteqn}{\setcounter{equation}{\value{saveeqn}}%
  \renewcommand{\theequation}{\Alph{subsection}.\arabic{equation}}%
  \renewcommand{\)}{\end{equation}}}

\renewcommand{\thefootnote}{\alph{footnote}}
\renewcommand{\(}{\begin{equation}}
\renewcommand{\)}{\end{equation}}
\newcommand{\ba}{\begin{eqnarray}}
\newcommand{\ea}{\end{eqnarray}}
\renewcommand{\a}{\alpha}

\renewcommand{\sl}{{\sqrt{\lambda}}}

\newcommand{\cbp}{\mathop{\vtop{\ialign{##\crcr
   $\hfil\displaystyle{}\hfil$\crcr\noalign{\kern-13pt\nointerlineskip}
   \BIG{)}\hskip0pt\crcr\noalign{\kern3pt}}}}}
\newcommand{\pa}{\mathop{\vtop{\ialign{##\crcr

$\hfil\displaystyle{\oplus}\hfil$\crcr\noalign{\kern+1pt\nointerlineskip
}
   \hspace{.08in}$^{\alpha=0}$\hskip6pt\crcr\noalign{\kern3pt}}}}}
\renewcommand{\hsp}{,\hspace{.3in}}
\newcommand{\p}{^\prime}
\newcommand{\pp}{^{\prime\prime}}

\catcode`\@=11
\def\vereq#1#2{\lower3pt\vbox{\baselineskip1.5pt \lineskip1.5pt
\ialign{$\m@th#1\hfill##\hfil$\crcr#2\crcr\sim\crcr}}}
\catcode`\@=12

\renewcommand{\(}{\begin{equation}}
\renewcommand{\)}{\end{equation}}

\def\vx{{\vec{x}}}

\def\vk{{\vec{k}}}

\def\k#1{\k_{#1}}

\def\pxink{\int \frac{dk_x}{2\pi}}
\def\pyink{\int \frac{dk_y}{2\pi}}

\def\pinvk{\int \frac{d^{2}\vk}{(2\pi)^{2}}}

\def\npinvk#1{\int \frac{d^{2}\vk_#1}{(2\pi)^{2}}}

\def\sinvk{\int\hspace{-17pt}\sum \frac{d^{2}\vk}{(2\pi)^{2}}}

\def\df{\mathcal{D}_{f}}

\def\omvk{\omega_{\vec{k}}}

\def\nomvk#1{\omega_{\vec{k}_{#1}}}

\def\bdvk{B^{\ddagger}_{\vec{k}}}
\def\bvk{B_{\vec{k}}}

\def\bvkm{B_{-\vec{k}}}

\def\nbdvk#1{B^{\ddagger}_{\vec{k}_{#1}}}

\def\g{\mathfrak g}
\def\gvk{\g_{\vk}}

\def\ngvk#1{\g_{\vk_{#1}}}

\usepackage[dvipsnames]{xcolor}

\newcommand{\beas}{\begin{eqnarray*}}
\newcommand{\eeas}{\end{eqnarray*}}

\newcommand{\bquo}{\begin{quote}}
\newcommand{\enqu}{\end{quote}}

\def\lim#1{\stackrel{\rm{lim}}{{}_{#1}}}

\def\ok#1{\omega_{k_{#1}}}

\def\V#1{V^{(#1)}(\sqrt{\lambda}f(x))}

\newcommand{\beq}{\begin{equation}}
\newcommand{\eeq}{\end{equation}}
\newcommand{\bea}{\begin{eqnarray}}
\newcommand{\eea}{\end{eqnarray}}

\newskip\humongous \humongous=0pt plus 1000pt minus 1000pt

\newif\ifdtup

\catcode`\@=11

\ifthenelse{\equal{\letter}{0}}{ 

\@addtoreset{equation}{section}
\def\theequation{\arabic{section}.\arabic{equation}}

\def\@normalsize{\@setsize\normalsize{15pt}\xiipt\@xiipt
\abovedisplayskip 14pt plus3pt minus3pt%
\belowdisplayskip \abovedisplayskip
\abovedisplayshortskip \z@ plus3pt%
\belowdisplayshortskip 7pt plus3.5pt minus0pt}

\def\small{\@setsize\small{13.6pt}\xipt\@xipt
\abovedisplayskip 13pt plus3pt minus3pt%
\belowdisplayskip \abovedisplayskip
\abovedisplayshortskip \z@ plus3pt%
\belowdisplayshortskip 7pt plus3.5pt minus0pt
\def\@listi{\parsep 4.5pt plus 2pt minus 1pt
      \itemsep \parsep
      \topsep 9pt plus 3pt minus 3pt}}

\relax

\def\section{\@startsection{section}{1}{\z@}{3.5ex plus 1ex minus  .2ex}{2.3ex plus .2ex}{\large\bf}}

\def\thesection{\arabic{section}}
\def\thesubsection{\arabic{section}.\arabic{subsection}}

\def\appendix{\setcounter{section}{0}
 \def\thesection{Appendix \Alph{section}}
 \def\thesubsection{\Alph{section}.\arabic{subsection}}
 \def\theequation{\Alph{section}.\arabic{equation}}}
\renewcommand{\theequation}{\arabic{section}.\arabic{equation}}

}{
\renewcommand{\theequation}{\arabic{equation}}

} 

\begin{document}
\def\thefootnote{\fnsymbol{footnote}}
\def\thetitle{The Domain Wall String's Anti-Stokes Scattering Cross Section}
\def\autone{Hengyuan Guo}
\def\auttwo{Jarah Evslin}
\def\autthree{Hui Liu}

\def\affa{School of Physics and Astronomy, Sun Yat-sen University, Zhuhai 519082, China}
\def\affaa{Department of Physics, ``Enrico Fermi`'',    University of Pisa,  \\
Largo Pontecorvo, 3, 56127, Pisa, Italy}
\def\affaaa{INFN, Sezione di Pisa,  
Largo Pontecorvo, 3, 56127, Pisa, Italy}
\def\affaaaa{ Lanzhou Center for Theoretical Physics,\\
Key Laboratory of Theoretical Physics of Gansu Province,\\
\\Key Laboratory of Quantum Theory and Applications of MoE, Lanzhou University, Lanzhou, Gansu 730000, China}
\def\affb{Institute of Modern Physics, NanChangLu 509, Lanzhou 730000, China}
\def\affc{University of the Chinese Academy of Sciences, YuQuanLu 19A, Beijing 100049, China}
\def\affd{Yerevan Physics Institute, 2 Alikhanyan Brothers St., Yerevan 0036, Armenia}


\ifthenelse{\equal{\pr}{1}}{
\title{\thetitle}
\author{\autone}
\author{\auttwo}
\author{\autthree}
\affiliation {\affa}
\affiliation {\affb}
\affiliation {\affc}

}{}

\begin{center}
{\large {\bf \thetitle}}

\bigskip

\bigskip


{\large \noindent  
\autone{${}^{1,2,3,4}$} 
\footnote{guohy57@mail.sysu.edu.cn},
 \auttwo{${}^{5,6}$} 
\footnote{jarah@impcas.ac.cn}
and
\autthree{${}^{7}$}
\footnote{hui.liu@yerphi.am}
 
}


\vskip.7cm

1) \affa\\
2) \affaa\\
3) \affaaa\\
4)  Lanzhou Center for Theoretical Physics,\\
Key Laboratory of Theoretical Physics of Gansu Province,\\
Key Laboratory of Quantum Theory and Applications of MoE, Lanzhou University, Lanzhou, Gansu 730000, China\\
5) \affb\\
6) \affc\\
7) \affd\\

\end{center}


\begin{abstract}
\noindent
 We consider anti-Stokes scattering, in which a perturbative meson scatters off of a domain wall string's shape mode excitation, de-exciting it.  Previously the probability of this process was calculated for a perpendicular incident meson striking the center of a localized shape mode.  The answer depended on the profiles of the initial wave packets.  In this paper we consider
 an arbitrary incident angle and impact parameter.  In addition to the de-excitation probability, we compute the cross section for this process, which as usual is independent of the details of the wave packets.  To our knowledge, this is the first time that a cross section has been defined for the scattering of a bulk degree of freedom with a localized excitation in an extended soliton.

\end{abstract}

%
\setcounter{footnote}{0}
\renewcommand{\thefootnote}{\arabic{footnote}}

\ifthenelse{\equal{\pr}{1}}
{
\maketitle
}{}

\section{Introduction}

Soliton solutions in weakly coupled classical field theories necessarily involve inverse powers of the coupling constant.  As a result, the states corresponding to solitons in quantum field theory are necessarily nonperturbative \cite{skyrme}.  In addition, in translation-invariant models, classical solitons break the translation symmetry and so possess a zero mode, which significantly complicates their quantization \cite{fkk75,fk76,mat77}.  Nonetheless, many formalisms have been developed over the years to treat solitons in quantum field theory \cite{sol1,sol2,sol3,sol4}.

Of these formalisms, the most efficient can only be applied at the one-loop level \cite{cq1,cq2,gw22}.  The more powerful formalisms, as a result of the issues above, have historically been quite complicated.  For example, the most popular approaches, the collective coordinate approaches of Refs.~\cite{gs74,cl75,tom75}, introduce the zero mode as a collective coordinate.  This makes the translation-invariance manifest, but the collective coordinate is not a canonical coordinate and so the quantization of the theory is quite involved.  A complicated canonical transformation is required, which adds an infinity of terms into the Hamiltonian.  It was later discovered that this transformation does not preserve the path integral, leading to another infinity of terms \cite{gj76}.  In addition, their application is complicated by the multivalued nature of the map from the local coordinates about a saddle point to the full field space~\cite{bh24}.  The result of these problems and others is that applications of such methods in the literature often lead to incorrect results, as described in Ref.~\cite{ga24}.  Moreover, phenomenological quantities such as the kink-meson scattering amplitude have only been calculated with an incomplete subset of contributions even at leading order \cite{hayashi1,hayashi2}.

This situation has changed in Refs.~\cite{mekink,me2loop} with the introduction of Linearized Soliton Perturbation Theory (LSPT).  Here manifest translation symmetry is sacrificed, and one instead directly quantizes the canonical coordinates corresponding to a decomposition of the fields linearized about the classical solution.  The advantage is that calculations become so simple that scattering amplitudes and probabilities have been calculated for several kink-meson scattering processes in 1+1 dimensions \cite{memult,mestokes}.

In 1+1 dimensions, the fundamental question is as follows.  A kink and meson begin moving towards one another, and one asks the probability of each final state.  This problem has been treated exhaustively at the leading orders.

Recently, Ref.~\cite{hengyuanstokes} has moved up to domain wall string-meson scattering in 2+1 dimensions.  One process considered was anti-Stokes scattering \cite{stokes52,raman28,landsberg28}, in which the string begins with a localized shape mode excitation, and a meson is fired into the middle of this excitation, with an initial velocity perpendicular to the string.  

The resulting probability, needless to say, depends on the shapes of the initial meson and shape mode wave packets and on the smearing of these wave packets during propagation.  This is because scattering in 2+1 dimensions, unlike 1+1 dimensions, involves an impact parameter.  As is well known, beyond 1+1 dimensions one is usually interested in the scattering cross section instead of the probability, because the cross section is independent of the details of the initial wave packets and their propagation.  Of course, in the case of kinks and more generally domain walls, one must work with localized wave packets and not plane waves because the two sides of the domain wall are in different vacua and so mesons in the two behave differently.  Nonetheless, the cross section only depends on the choice of vacuum from which the meson is fired, and not on the shape of its wave packet.

With this motivation, in the present paper we again treat anti-Stokes scattering, but our aim is to calculate the cross section instead of the probability.  As we are interested in the cross section, we will ignore the wave packet spreading which we know will not affect our answer.  However, unlike Ref.~\cite{hengyuanstokes} we do allow for an arbitrary incident angle of the meson and also a longitudinal offset between the initial meson position and the shape mode position.  We use the resulting impact parameter dependence of the probability to define a notion of cross section which is suitable for the scattering of bulk mesons with localized excitations inside of the domain wall string.  We analytically evaluate the cross section in a scalar model with a fairly arbitrary potential and our result is quite simple.

What about Stokes scattering?  This is the process in which a meson strikes an unexcited domain wall string, exciting it.  In this case the entire domain wall is the target, not a localized excitation.  Intuitively, the trajectory of the meson is one-dimensional and it strikes a codimension one target, therefore the occurence of a collision is independent of continuous changes in parameters such as the wave function shape.  This indeed was the conclusion of the calculation in Ref.~\cite{hengyuanstokes}, the Stokes scattering probability is independent of the shapes of the various initial wave functions.  In that case, probability is an appropriate quantity to characterize the scattering and there is no sense in introducing a cross section.  Indeed, the cross section for Stokes scattering with an infinite domain wall string would be infinite.

Similarly to Ref.~\cite{hengyuanstokes}, the motivation for this study arises from cosmology.  In Ref.~\cite{1703.06696}, it was found that simulations of cosmic strings are poorly described by the Nambu-Goto action when the strings interact strongly with radiation.  In particular, the uniquitous small loops found in Nambu-Goto simulations are absent in the full field theory simulations.  In Ref.~\cite{0812.1929}, it was speculated that this is a result of the small scale structure of strings, and in particular vibrations on the scale of their width.  

A tantalizing clue in this direction was recently reported in Ref.~\cite{2405.06030}, which found radiation at precisely twice the frequency of the internal shape mode.  This motivates a study of the interactions of string shape modes and radiation.  In classical field theory, Refs.~\cite{2209.12945,2411.13521} have initiated a program of studying such interactions in the case of the domain wall string in 2+1 dimensional models.  The current work is an extension of that program to quantum field theory.

Needless to say, the domain wall string is not only a model of a cosmic string but also a domain wall.  Very recently, in Ref.~\cite{ai25} it has been shown that in cosmological first order phase transitions, quantum interactions of radiation fields with domain walls can have a qualitatively important effect on their velocities.  This further motivates an understanding of quantum interactions between radiation and domain walls.

We begin in Sec.~\ref{revsez} with a review of LSPT, the formalism that will be used.  Our general formula for the anti-Stokes scattering cross section is derived in Sec.~\ref{asez}.  Finally in Sec.~\ref{fsez} we apply this general result to the example of the $\phi^4$ double-well model.

\section{Linearized Soliton Perturbation Theory} \label{revsez}

Our calculation of the anti-Stokes scattering cross section will use a formalism called Linearized Soliton Perturbation Theory (LSPT), developed in Refs.~\cite{mekink,me2loop}.  In the present section, we shall review the relevant features of this formalism.

\subsection{Decomposing the States}

Consider a classical field theory in two spatial dimension $\vx=(x,y)$ with a single scalar field $\phi(\vx,t)$ and its conjugate momentum $\pi(\vx,t)$.  The model is described by a Hamiltonian $H[\phi,\pi]$.  Imagine that there is a stationary solution
\beq
\phi(\vx,t)=f(x)\hsp \pi(\vx,t)=0 \label{cl}
\eeq
which is independent of the $y$ direction.  We will be interested in the case in which the solution $f(x)$ corresponds to a domain wall.

We may promote this classical field theory to a quantum field theory.  We will write the Schrodinger picture quantum field as $\phi(\vx)$ and its conjugate momentum as $\pi(\vx)$.  

Now we pose the question: What state $|K\rangle$ in the quantum field theory corresponds to our classical solution (\ref{cl})?  We would like a stationary state, and so it should be a Hamiltonian eigenstate
\beq
H|K\rangle=Q|K\rangle. \label{heq}
\eeq
It has long been appreciated \cite{vinc72,cornwall74} that the answer is some kind of coherent state.  It is also known that, beyond 1+1 dimensions, coherent states lead to divergences in the energy density \cite{erice} and other quantities \cite{cocorr23}.  

The starting point of LSPT is that such a deformed coherent state may be written
\beq
|K\rangle=\df \vac\hsp
\df={{\rm Exp}}\left[-i\int d^2\vx f(x)\pi(\vx)\right].
\eeq
Here $\df$ is the displacement operator, which shifts the field by the classical solution $f(x)$.  This decomposition factorizes our state $|K\rangle$ into a part $\df$ which is nonperturbative but already known and a part $\vac$ which we will show can be found using perturbative techniques.

In fact, any state $|\Psi\rangle$ consisting of a domain wall plus a finite number of bound excitations and unbound perturbative quanta, which we will call mesons, can be decomposed in this way 
\beq
|\Psi\rangle=\df|\psi\rangle
\eeq
where $|\psi\rangle$ lives in an ordinary, perturbative Fock space.  Amplitudes will be inner products of such states.  To compute such inner products, one can forget about $\df$ and the original $|\Psi\rangle$ because
\beq
\langle \Psi_1|\Psi_2\rangle=\langle \psi_1|\psi_2\rangle.
\eeq
In other words, we only need to understand the states $|\psi\rangle$ and their evolution in order to calculate observables.

\subsection{Reformulating the Spectral and Time Evolution Problems}

So what are the states $|\psi\rangle$?  It is easy to see that our original eigenvalue problem
\beq
H|\Psi\rangle=E|\Psi\rangle
\eeq
is equivalent to the eigenvalue problem
\beq
H\p|\psi\rangle=E|\psi\rangle\hsp
H\p=\df^\dag H\df
\eeq
where we call $H\p$ the domain wall Hamiltonian.  Furthermore, evolving the Schrodinger picture state $|\Psi\rangle$ by a time $t$, one finds
\beq
e^{-iHt}\df|\psi\rangle=\df e^{-iH\p t}|\psi\rangle.
\eeq
We learn that the spectrum of the operator $H\p$ consists of the states $|\psi\rangle$ and also that $H\p$ acts as the time evolution operator on these states.

We have reformulated the spectral and time evolution problems from problems in terms of $H$ to problems in terms of $H\p$.  What have we gained?  While the operator $H\p$ is no simpler than $H$, the states $|\psi\rangle$ on which it acts are much simpler than the old states $|\Psi\rangle$.  This is because the old states $|\Psi\rangle$ contain a $\df$ operator, which in the case of a soliton such as a domain wall is an exponential of an expression with an inverse power of the coupling, and so is hopelessly nonperturbative.  Therefore we have converted a hopelessly nonperturbative problem into a problem that we may be able to solve in perturbation theory.

\subsection{The Domain Wall Hamiltonian $H\p$}

Can we solve these problems in perturbation theory?  Of course this depends on our Hamiltonian $H$.  Let us consider
\begin{equation}
H=\int d^2\vx :\mathcal{H}(\vx): \hsp
\mathcal{H}(\vx)=\frac{\pi^2(\vx)}{2}+\frac{(\partial_i \phi(\vx))^2}{2}
+\frac{V(\sqrt{\lambda} \phi(\vx))}{\lambda}
\end{equation}
where $::$ is the usual normal ordering.  This normal ordering removes all diagrams with a loop based at a single vertex, which are the most divergent.

Here $\lambda$ is a coupling constant, which will be taken to be small so that we may perform a semiclassical expansion.  $V$ is a potential and, in order to have a domain wall solution $f(x)$, it will be required to have degenerate minima $\phi_1$ and $\phi_2$ so that $f(-\infty)=\phi_1$ and $f(\infty)=\phi_2$.  We will define the mass $m$ of our scalar field by
\beq
m^2=V^{(2)}(\sqrt{\lambda} \phi_1)=V^{(2)}(\sqrt{\lambda} \phi_2)\hsp
V^{(n)}(\sqrt{\lambda} \phi(\vx))=\frac{\partial^n V(\sqrt{\lambda} \phi(\vx))}{(\partial \sqrt{\lambda} \phi(\vx))^n}
\eeq
so that the masses in the two vacua agree.  If we did not impose this condition, our domain wall would accelerate in the quantum theory \cite{wstabile}.

The displacement operator $\df$ shifts the Schrodinger picture field by the classical solution and is compatible with normal ordering, so that for any functional $F$
\beq
:F(\phi,\pi):\df=\df :F(\phi+f,\pi):.
\eeq
In the case of the Hamiltonian, this implies
\beq
H\p[\phi,\pi]=\df^\dag H[\phi,\pi]\df=H[\phi+f,\pi]. \label{hpd}
\eeq
Expanding $H\p$ in powers of the coupling
\beq
H\p=\sum_i H\p_i\hsp H\p_i\sim O(\lambda^{-1+i/2})
\eeq
Eq.~(\ref{hpd}) yields the domain wall Hamiltonian
\bea
H\p_0&=&Q_0\hsp H\p_1=0\\
H\p_2&=&\frac{1}{2}\int d^2\vx\left[:\pi^2(\vx):+:\left(\partial_x\phi(\vx)\right)^2:+:\left(\partial_y\phi(\vx)\right)^2:+\V2:\phi^2(\vx):\right.]\nonumber\\
H\p_{n>2}&=&\lambda^{\frac{n}{2}-1}\int d^2\vx \frac{V^{(n)}(\sqrt{\lambda} f(x))}{n !}: \phi^n(\vx):\nonumber
\eea
where $Q_0$ is the energy of the classical domain wall.  We are interested in an infinite domain wall, and so this is infinite, but it is the $y$ integral of the finite domain wall tension $\rho_0$.

\subsection{Normal Modes}

Can we solve the $H\p$ eigenvalue problem perturbatively?  To do this, we need to exactly diagonalize $H\p$ at all orders up to $O(\lambda^0)$, producing an approximation $|\psi\rangle_0$  to the state $|\psi\rangle$.  Then, we can use $H\p_{n>2}$ to find perturbative corrections to $|\psi\rangle_0$ using old fashioned perturbation theory.  For leading order calculations, like the tree-level scattering that we will compute, we will not need these perturbative corrections.

So how do we find $|\psi\rangle_0$?  $H\p$ contains three components up to order $O(\lambda^0)$.  First there is $H\p_0$.  This is of order $O(1/\lambda)$ and so in principle could ruin our perturbative approach, however it is a $c$-number and so is diagonalized by any state.  The next term, $H\p_1$, is of order $O(1/\sl)$, however so long as $f(x)$ satisfies the classical equations of motion it vanishes. 

We conclude that we only need to diagonalize a single operator $H\p_2$.  We recognize it as the Hamiltonian for the free massive scalar theory, which is ordinarily easy to diagonalize using the operators $a^\dag_p$ and $a_p$ that create and destroy plane waves.  However, in this case, the mass squared $\V2$ depends on $x$.  This means that in order to diagonalize it, we need to Bogoliubov transform the operators $a^\dag_p$ and $a_p$ into some new operators that create and destroy constant frequency perturbations $\g(x)$ of the solution $f(x)$, as described in Refs.~\cite{wentzel,cahill76} in 1+1 dimensions and in Ref.~\cite{noi21} in the present context.

More precisely, one begins with a constant frequency $\omega$ perturbation to our classical solution
\beq
\phi(\vx,t)=f(x)+\g_{k_x}(x)e^{-ik_y y-i\omega_{k_xk_y} t}\hsp \omega_{k_xk_y}\geq 0. \label{an}
\eeq
Here $k_y$ is a real number, the $y$ momentum, while $k_x$ is an abstract index that will be described shortly.  It will be convenient to assemble $k_x$ and $k_y$ into a vector
\beq
\vk=(k_x,k_y)\hsp \g_{\vk}(\vx)=\g_{k_x}(x)e^{-ik_y y}.
\eeq

One then substitutes (\ref{an}) into the classical equations of motion and drops terms of order $O(\g^2)$.  This leads to the Sturm-Liouville equation
\beq
\V{2}\g_{k_x}(x)=\omega_{k_x}^2\g_{k_x}(x)+\g\pp_{k_x}(x) \hsp \omega_{k_x}=\sqrt{\omega_{k_xk_y}^2-k_y^2}.
\eeq
Its solutions, which are the normal modes, generate the space of functions on the real numbers and so, together with $e^{-ik_yy}$ can be used to expand our Schrodinger picture fields $\phi(\vx)$ and $\pi(\vx)$.

The solutions can be classified according to the value of the frequency $\ok{x}$.  First, there is always a zero mode
\beq
\g_B(x)=-\frac{f\p(x)}{\sqrt{\rho_0}}\hsp \omega_B=0.
\eeq
In this case, the index label $k_x$ is the letter $B$.
Next, for every $\omega_{k_x}>m$, there are two solutions.  We will name the corresponding $k_x$ using the convention $ k_x=\pm\sqrt{\omega^2_{k_x}-m^2}$.  In the case of a reflectionless potential, they are asymptotically proportional to $e^{-ik_x x}$.  Finally, there may or may not be discrete shape modes $\g_S(x)$ with frequencies $0<\omega_S<m$.  Overall we conclude that the abstract index $k_x$ runs over discrete indices $\{B,S_1,\cdots,S_n\}$ corresponding to localized solutions and also all real numbers corresponding to continuum solutions.

We adopt the following conventions.  First, bound modes are taken to be real and they satisfy
\beq
\omega_{Bk_y}=|k_y|\hsp \omega_{Sk_y}=\sqrt{\omega_{S}^2+k_y^2}\hsp \int dx \g^2_{B}(x)=1\hsp \int dx \g_{S}(x)\g_{S\p}(x)=\delta_{SS\p}.
\eeq
The continuum modes satisfy
\beq
\omega_{k_xk_y}=\sqrt{m^2+k_x^2+k_y^2}\hsp 
\g^*_{\vk}(\vx)=\g_{-\vk}(\vx)\hsp
\int dx \g_{k_x}(x)\g^*_{k\p_x}(x)=2\pi\delta(k_x-k\p_x).
\eeq
In our abstract vector notation $\vk=(k_x,k_y)$ the shape mode frequencies above would be written
\beq
\vk=(S,k_y)\hsp \omega_\vk=\sqrt{\omega_S^2+k_y^2}.
\eeq

Finally, we will decompose\footnote{Strictly speaking we should include the case $\vk=(B,0)$ separately, as $\omega_\vk$ vanishes in that case.  However it will play no role in the present paper.} the fields in terms of the normal modes as
\beq\label{dec}
\phi(\vx) =\sinvk\left(\bdvk+\frac{\bvkm}{2 \omvk}\right)\g_{\vk}(\vx) \hsp
\pi(\vx) =i \sinvk\left(\omvk \bdvk-\frac{\bvkm}{2}\right)\gvk(\vx)
\eeq
where the integral is defined as
\beq
\sinvk=\pyink\left(\pxink+\sum_{B,S}\right).
\eeq
The sum is a sum over all discrete values of $k_x$.  We have defined the annihilation operators $B_\vk$ and the creation operators
\beq
B^\ddag_\vk=\frac{B^\dag_\vk}{2\omega_\vk}
\eeq
with the conventions
\beq
-(B,k_y)=(B,-k_y)\hsp -(S,k_y)=(S,-k_y).
\eeq

The canonical commutation relations satisfied by $\phi(\vx)$ and $\pi(\vx)$ imply that the creation and annihilation operators satisfy
\bea
\left[B_{Bk_y},
B_{B k_y\p}^{\ddagger}\right]&=&2\pi\delta(k_y-k_y\p)\\
\left[B_{S k_y},B_{S\p k_y\p}^{\ddagger}\right]&=&\delta_{S S\p}2\pi\delta(k_y-k_y\p)
\hsp\left[B_{\vk_1}, B_{\vk_2}^{\ddagger}\right]=(2\pi)^2\delta^2\left(\vk_1-\vk_2\right).\nonumber
\eea

\subsection{The Domain Wall Sector}

We have introduced all of this notation because of the following simplification \cite{cahill76,noi21}
\beq
H\p_2=\int dy \rho_1 +\int\frac{dk_y}{2\pi}\left(|k_y|{B_{Bk_y}^{\ddag} B_{Bk_y}}+\sum_S\omega_{Sk_y} B_{Sk_y}^{\ddag} B_{Sk_y}\right)+\pinvk\omvk\bdvk\bvk.
\eeq
Here $\rho_1$ is the one-loop correction to the tension of the domain wall string.  We see that $H\p_2$ is a sum of a $c$-number term and a set of quantum harmonic oscillators, one for each normal mode.  Therefore, one may immediately write down its spectrum.

The ground state of $H\p_2$ is $\vac_0$, defined to be the state in which all normal modes are switched off
\beq
B_{Bk_y}\vac_0=B_{Sk_y}\vac_0=\bvk\vac_0=0. \label{v0}
\eeq
As the $B$ oscillators are related to the $a$ operators by a linear Bogoliubov transformation, we recognize $\vac_0$ as a squeezed state and the corresponding $\df\vac_0$ approximation to the domain wall ground state as a squeezed coherent state.

The excited states can be found by acting on $\vac_0$ with the creation operators.  Below we will be interested in two excited states.  The first is a state consisting of a ground state domain wall and a perturbative meson
\beq
|\vk\rangle_0=B^\ddag_\vk\vac_0.
\eeq
Here $\vk=(k_x,k_y)$ where $k_x$ is a real number.  If the meson is folded into a nearly monochromatic wave packet localized far from the domain wall, then $\vk$ will approximate its momentum.  Second, we will be interested in a state that again has a perturbative meson, but now the domain wall has an excited shape mode moving with momentum $k_{Sy}$
\beq
|\vk_S\vk\rangle_0=B^\ddag_{\vk_S}B^\ddag_{\vk}\vac_0\hsp \vk_S=(S,k_{Sy}).
\eeq

\section{Anti-Stokes Scattering} \label{asez}

In anti-Stokes scattering, a meson wave packet is incident on an excited domain wall string.  The domain wall de-excites and then re-emits the meson.  We let $\vk_1$ be the initial meson momentum.  The domain wall's excited shape mode begins in a wave packet centered at the origin with a momentum of $k_{Sy}$ parallel to the wall, which we again describe using the abstract vector
\beq
\vk_S=(S,k_{Sy}).
\eeq

\subsection{Initial Condition}

The initial state is thus
\beq
\left|\Phi\right\rangle_0=
\npinvk1\a_{\vk_1}\int\frac{dk_{Sy}}{2\pi}e^{-\sigma_0^{2}k_{Sy}^2}\left|\vk_S \vk_1\right\rangle_0\hsp
|\vk_S \vk_1\rangle_0=B^\ddag_{Sk_{Sy}} \nbdvk1\vac_0\label{inita}
\eeq
where $\alpha_{k_1}$ is the momentum-space wave function of the initial meson.

This process is somewhat different from Stokes scattering because the incoming meson may miss the localized shape mode, in which case there will be no reaction.  More concretely, when the meson reaches the wall, it will be separated from the shape mode in the $y$ direction by some impact parameter $b$.  If $b$ is greater than the widths of the meson and shape mode wave packets,  then we expect the amplitude for this interaction to be suppressed.  By integrating over different values of the impact parameter, we could determine the cross section of the shape mode for this process.

Ref.~\cite{hengyuanstokes} considered the case in which $\sigma_0$ is so large that, given the initial meson position $(x_0,y_0)$, the meson necessarily passes through the shape mode.  Here we will make no such assumption, and indeed to find the cross section we will need to understand the role of the impact parameter.  
Also, while the study \cite{hengyuanstokes} considered a meson moving perpendicular to the string, we will instead allow an arbitrary tangential momentum $k_{0y}$.  In all, we consider the initial condition
\bea 
\Phi(\vx)&=&e^{-\frac{(x-x_0)^2}{4\sigma^2}+ix k_{0x}}e^{-\frac{(y-y_0)^2}{4\sigma^2}+iy k_{0y}}\hsp x_0<0\hsp k_{0x}>0  \label{aspach}\\
\alpha_{\vk_1}&=&\int d^2\vx \Phi(\vx)\ngvk1(\vx)
\hsp  \left\{\frac{1}{k_{0x}},\frac{1}{m}\right\} \ll \{\sigma,\sigma_0\} \ll |x_0|.  \nonumber
\eea
This describes a meson that begins in a wave packet of diameter $\sigma$ localized at $(x_0,y_0)$ and moving with momentum $\vk_0=(k_{0x},k_{0y})$ and also a shape mode that begins that at rest, at the origin, in a wave packet of width $\sigma_0$.



In Refs.~\cite{mestokes,hengyuanstokes} we have seen that the generalization of a scattering amplitude from a reflectionless to a generic potential is very simple.  Therefore, to avoid clutter, we will consider a model with a reflectionless potential $V$.  As a result, far from the wall the normal modes reduce to plane waves and so our initial meson wave packet in momentum space is simply the Gaussian
\beq
\begin{aligned}
  \alpha_{\vk_1}=&\int d^2\vx \Phi(\vx)\ngvk1=\int d^2\vx \Phi(\vx)e^{-ik_{1x}x-ik_{1y}y}\\
=&\int d^2\vx e^{-\frac{(x-x_0)^2}{4\sigma^2}+ix( k_{0x}-k_{1x})}e^{-\frac{(y-y_0)^2}{4\sigma^2}+iy (k_{0y}-k_{1y})} \\
=&(2\sigma\sqrt{\pi})^2e^{-\sigma^2(k_{0x}-k_{1x})^2+ix_0(k_{0x}-k_{1x})}
e^{-\sigma^2(k_{0y}-k_{1y})^2+iy_0(k_{0y}-k_{1y})}. \label{reflectionlessa}
\end{aligned}
\eeq

\subsection{Evolving the Initial State}

At $O(\sl)$, as shown in the Appendix of Ref.~\cite{hengyuanstokes}, the only term in the Hamiltonian $H\p$ which can transform a one-meson, one excited string state into a one-meson, one de-excited string state is
\beq
\begin{aligned}
   H_I=&\frac{\sqrt{\lambda}}{4}\int\frac{dk_{Sy}}{2\pi}\npinvk1\npinvk2\frac{V_{\vk_2,-\vk_S,-\vk_1}}{\omega_{\vk_S} \nomvk1} \nbdvk2 B_{\vk_S} B_{\vk_1}  \\
H_I |\vk_{S} \vk_1\rangle_0=&\frac{\sqrt{\lambda}}{4}\npinvk2\frac{V_{\vk_2,-\vk_S,-\vk_1}}{\nomvk1\omega_{\vk_{S}}}\left|\vk_2\right\rangle_0 \label{hi}\\ 
\end{aligned}
\eeq
where we have defined the three-point vertex
\bea
V_{\vk_2,-\vk_S,-\vk_1}&=&\int d^2\vx V^{(3)}(\sqrt{\lambda} f(x))\g_{S,-k_{Sy}}(\vx)\g_{\vk_2}(\vx)\g_{-\vk_1}(\vx)=V_{S,k_{2x},-k_{1x}}(2\pi)\delta(-k_{Sy}+k_{2y}-k_{1y})\nonumber\\
V_{S,k_2,-k_1}&=&\int dx V^{(3)}(\sqrt{\lambda}f(x))g_S(x)g_{k_2}(x)g_{-k_1}(x). \nonumber
\eea

The corresponding leading order contribution to an evolved one-meson, one excited string basis state $|\vk_S\vk_1\rangle_0$ is
\beq
\begin{aligned}
 e^{-iH\p t}|\vk_S\vk_1\rangle_0\bigg|_{O(\sl)}
 &=-i\int_0^t d\tau e^{-i(t-\tau )H_2\p}H_I e^{-iH_2\p\tau}|\vk_S\vk_1\rangle_0 \label{eva}
 =-i\int_0^t d\tau e^{-i(t-\tau)H_2\p} H_I e^{-i\tau (\nomvk1+\omega_{\vk_S})}|\vk_S\vk_1\rangle_0\\
=&\frac{-i\sqrt{\lambda}}{4}  \npinvk2 \int_0^t d\tau e^{-i(t-\tau)H_2\p}e^{-i\tau (\nomvk1+\omega_{\vk_S})}\frac{V_{\vk_2,-\vk_S,-\vk_1}}{\nomvk1\omega_{\vk_S}}|\vk_2\rangle_0\\
=&\frac{-i\sqrt{\lambda}}{4}\npinvk2\int_0^t d\tau e^{-i(t-\tau)\nomvk2}e^{-i\tau (\nomvk1+\omega_{\vk_S})}\frac{V_{\vk_2,-\vk_S,-\vk_1}}{\nomvk1\omega_{\vk_S}}|\vk_2\rangle_0\\
=&\frac{-i\sqrt{\lambda}}{4} \npinvk2 \int_0^t d\tau e^{-it\nomvk2}e^{-i\tau (\nomvk1+\omega_{\vk_S}-\nomvk2)}\frac{V_{\vk_2,-\vk_S,-\vk_1}}{\nomvk1\omega_{\vk_S}}|\vk_2\rangle_0.
\end{aligned}
\eeq

Folding this into the initial state (\ref{reflectionlessa}), one obtains our master formula for the state at a time $t$
\beq
\begin{aligned}
 &e^{-iH\p t}|\Phi\rangle_0\bigg|_{O(\sl)}\\
 =&\frac{-i\sqrt{\lambda}}{4}\int\frac{dk_{Sy}}{2\pi}e^{-\sigma_0^{2}k_{Sy}^2}\npinvk1 
\frac{\a_{\vk_1}}{\nomvk1}\npinvk2 \int_0^t d\tau e^{-it\nomvk2}e^{-i\tau (\nomvk1+\omega_{\vec{S}}-\nomvk2)}\frac{V_{\vk_2,-\vk_S,-\vk_1}}{\nomvk1\omega_{\vk_S}}|\vk_2\rangle_0\\
=&\int\frac{dk_{Sy}}{2\pi}\frac{-i\sqrt{\lambda}e^{-\sigma_0^{2}k_{Sy}^2}}{4\omega_{\vk_S}}\npinvk1 
\frac{(2\sigma \sqrt{\pi})^2 e^{-\sigma^2(k_{0x}-k_{1x})^2+ix_0(k_{0x}-k_{1x})}
e^{-\sigma^2(k_{0y}-k_{1y})^2+iy_0(k_{0y}-k_{1y})}}{\nomvk1}\\
&\times \npinvk2 \int_0^t d\tau e^{-it\nomvk2}e^{-i\tau (\nomvk1+\omega_{\vec{S}}-\nomvk2)}V_{\vk_2,-\vk_S,-\vk_1}|\vk_2\rangle_0\\
=&-i\pi\sigma^2\sqrt{\lambda}\npinvk2e^{-it\nomvk2}\int\frac{dk_{Sy}}{2\pi}e^{-\sigma_0^{2}k_{Sy}^2}\int_0^t d\tau \npinvk1 
\frac{V_{\vk_2,-\vk_S,-\vk_1}}{\nomvk1\omega_{\vec{k}_S}}J(\vk_1,\vk_2,k_{Sy},\tau)|\vk_2\rangle_0\\
&J(\vk_1,\vk_2,k_{Sy},\tau)=e^{-\sigma^2(k_{0x}-k_{1x})^2+ix_0(k_{0x}-k_{1x})-\sigma^2(k_{0y}-k_{1y})^2+iy_0(k_{0y}-k_{1y})-i\tau(\nomvk1+\omega_{\vk_S}-\nomvk2)}.
\end{aligned}
\eeq
Most of the computation will be the evaluation of these integrals.

Using the identity
\beq
V_{\vk_2,-\vk_S,-\vk_1}=V_{k_{2x},S,-k_{1x}}2\pi\delta(k_{2y}-k_{Sy}-k_{1y})
\eeq
to perform the $k_{Sy}$ integration, one arrives at
\bea
e^{-iH\p t}|\Phi\rangle_0\bigg|_{O(\sl)}&=&-i\pi\sigma^2\sqrt{\lambda}\npinvk2e^{-it\nomvk2}\\
&&\ \ \times
\int_0^t d\tau \npinvk1 
\frac{V_{k_{2x},S,-k_{1x}}}{\nomvk1\omega_{S,k_{2y}-k_{1y}}}e^{-\sigma_0^{2}(k_{2y}-k_{1y})^2}J(\vk_1,\vk_2,k_{2y}-k_{1y},\tau)|\vk_2\rangle_0\nonumber\\
J(\vk_1,\vk_2,k_{2y}-k_{1y},\tau)&=&e^{-\sigma^2(k_{0x}-k_{1x})^2+ix_0(k_{0x}-k_{1x})-\sigma^2(k_{0y}-k_{1y})^2+iy_0(k_{0y}-k_{1y})-i\tau(\nomvk1+\omega_{S,k_{2y}-k_{1y}}-\nomvk2)}
.\nonumber
\eea

Expanding $\nomvk1$ to first order about $\nomvk0$, one finds
\bea
\nomvk1&=&\nomvk0+\frac{\partial\nomvk1}{\partial k_{1x}}|_{\vk_0} (k_{1x}-k_{0x})+\frac{\partial\nomvk1}{\partial k_{1y}}|_{\vk_0}(k_{1y}-k_{0y})+O((k_{1y}-k_{0y})^2\label{omex}\\
&\sim& \nomvk0+\frac{k_{0x}}{\nomvk0}(k_{1x}-k_{0x})+\frac{k_{0y}}{\nomvk0}(k_{1y}-k_{0y})\nonumber\\
\omega_{S,k_{2y}-k_{1y}}&\sim&\omega_{S,k_{2y}-k_{0y}}+\frac{k_{0y}-k_{2y}}{\omega_{S,k_{2y}-k_{0y}}}(k_{1y}-k_{0y}).\nonumber
\eea
Let us define the positions
\beq
x_{\tau}=x_0+\frac{k_{0x}}{\nomvk0}\tau\hsp
y_{\tau}=y_0+\frac{k_{0y}}{\nomvk0}\tau\hsp 
y^S_{\tau}=\frac{k_{2y}-k_{0y}}{\omega_{S,k_{2y}-k_{0y}}}\tau.
\eeq
Then $J$ can be rewritten
\beq
J(\vk_1,\vk_2,k_{2y}-k_{1y},\tau)=e^{-\sigma^2(k_{0x}-k_{1x})^2+ix_{\tau}(k_{0x}-k_{1x})-\sigma^2(k_{0y}-k_{1y})^2+i(y_{\tau}-y^S_\tau)(k_{0y}-k_{1y})-i\tau(\nomvk0+\omega_{S,k_{2y}-k_{0y}}-\nomvk2)}
\eeq
and so the evolved state is
\bea
e^{-iH\p t}|\Phi\rangle_0\bigg|_{O(\sl)}&=&-i\pi\sigma^2\sqrt{\lambda}\npinvk2e^{-it\nomvk2}\int_0^t d\tau e^{-i\tau(\nomvk0+\omega_{S,k_{2y}-k_{0y}}-\nomvk2)}|\vk_2\rangle_0\\
&&\ \ \times
\npinvk1 
\frac{V_{k_{2x},S,-k_{1x}}}{\nomvk1\omega_{S,k_{2y}-k_{1y}}}e^{-\sigma^2(k_{0x}-k_{1x})^2-\sigma^2(k_{0y}-k_{1y})^2-\sigma_0^{2}(k_{2y}-k_{1y})^2}\nonumber\\
&&\ \ \ \ \ \ \times e^{ix_{\tau}(k_{0x}-k_{1x})+i(y_{\tau}-y_\tau^S)(k_{0y}-k_{1y})}.\nonumber
\eea

In the support of the Gaussian, we have argued that $\vk_1\sim \vk_0$.  We kept the linear correction in the exponential, as it is multiplied by a large time.  However, in the $\omega$ terms in the denominator, there is no such large time and so we may simply consider the constant term in the expansion.  The same is true for the interaction $V_{kSk}$ because it has support at $|x|\sim O(m)$ and so it varies slowly with respect to $k_{1x}$.  These approximations are exact in our large $\sigma$ limit, and lead to
\beq
e^{-iH\p t}|\Phi\rangle_0\bigg|_{O(\sl)}=-i\pi\sigma^2\sqrt{\lambda}\npinvk2e^{-it\nomvk2}\frac{V_{k_{2x},S,-k_{0x}}}{\nomvk0\omega_{S,k_{2y}-k_{0y}}}|\vk_2\rangle_0\int_0^t d\tau e^{-i\tau(\nomvk0+\omega_{S,k_{2y}-k_{0y}}-\nomvk2)} K
\eeq
where $K$ is 
\beq
K=\npinvk1 
e^{-\sigma^2(k_{0x}-k_{1x})^2-\sigma^2(k_{0y}-k_{1y})^2-\sigma_0^{2}(k_{2y}-k_{1y})^2+ix_{\tau}(k_{0x}-k_{1x})+i(y_{\tau}-y_\tau^S)(k_{0y}-k_{1y})}.
\eeq
The $\vk_1$ integrand in $K$ is now a Gaussian and so is easily evaluated.  Rewriting
\bea
\sigma^2(k_{0y}-k_{1y})^2+\sigma_0^{2}(k_{2y}-k_{1y})^2&=&\hat{\sigma}^2\left(k_{1y}-k_{0y}-\frac{\sigma_0^2}{\hat{\sigma}^2}(k_{2y}-k_{0y})\right)^2+D\nonumber\\
\hat{\sigma}\label{compeq}&=&\sqrt{\sigma^2+\sigma_0^2}\hsp D=\frac{\sigma_0^2\sigma^2(k_{0y}-k_{2y})^2}{\hat{\sigma}^2}
\eea
the integral is
\beq
K=\frac{1}{4\pi\sigma\hat{\sigma}}e^{
-\frac{x_\tau^2}{4\sigma^2}-\frac{(y_\tau-y_\tau^S)^2}{4\hat{\sigma}^2}-D-i \frac{\sigma_0^2}{\hat{\sigma}^2}(k_{2y}-k_{0y}) (y_\tau-y_\tau^S)
}.
\eeq
We conclude
\bea
e^{-iH\p t}|\Phi\rangle_0\bigg|_{O(\sl)}&=&-i\frac{\sigma}{4\hat{\sigma}}\sqrt{\lambda}\npinvk2e^{-D-it\nomvk2}\frac{V_{k_{2x},S,-k_{0x}}}{\nomvk0\omega_{S,k_{2y}-k_{0y}}} L|\vk_2\rangle_0\label{noas} \\
L&=&
\int_0^t d\tau e^{-\frac{x_\tau^2}{4\sigma^2}-\frac{(y_\tau-y_\tau^S)^2}{4\hat{\sigma}^2}}e^{
-i\tau(\nomvk0+\omega_{S,k_{2y}-k_{0y}}-\nomvk2)-i \frac{\sigma_0^2}{\hat{\sigma}^2}(k_{2y}-k_{0y}) (y_\tau-y_\tau^S)
}.\nonumber
\eea

Let us pause to interpret $L$.  This is an integral over the interaction time $\tau$.  The first Gaussian enforces that $\tau$ is the time that it takes the meson to reach the kink.  The second Gaussian is the condition that the shape mode is at the same $y$ coordinate as the meson when it arrives at the kink.  The first phase enforces the conservation of energy for an incoming meson of momentum $\vk_0$, which is the center of the momentum-space wave packet.

What about the last phase?  Recall from Eq.~(\ref{compeq}) that although the center of the incoming wave packet is indeed $\vk_0$, different parts of the wave packet are more likely to contribute to the final momentum $\vk_2$ and so the dominant contribution to the amplitude arises not from $\vk_1=\vk_0$ but rather from $k_{1y}=k_{0y}+(\sigma_0^2/\hat{\sigma}^2)(k_{2y}-k_{0y})$.  Of course this result is correct, but the fact that the dominant contribution is off-center of the wave packet complicates the calculation without adding any novel physics.  

We will therefore simplify the calculation by taking the limit $\sigma_0/\sigma\rightarrow 0$.  More precisely, we want $k_{1y}-k_{0y}$ to be much less than $1/\sigma$, so that $1/\sigma$ is a good approximation to the width of the momentum profile.  Therefore we consider
\beq
\frac{\sigma_0}{\sigma}\ll \frac{1}{\sigma_0 |k_{0y}|}.
\eeq
The right hand side is already much less than unity and so the initial shape mode wave packet is much more narrow than the initial meson wave packet.  

We then find
\bea
e^{-iH\p t}|\Phi\rangle_0\bigg|_{O(\sl)}&=&-\frac{i}{4}\sqrt{\lambda}\npinvk2e^{-\frac{\sigma_0^2\sigma^2}{\hat{\sigma}^2}(k_{0y}–k_{2y})^2-it\nomvk2}\frac{V_{k_{2x},S,-k_{0x}}}{\nomvk0\omega_{S,k_{2y}-k_{0y}}} L|\vk_2\rangle_0 \\
L&=&
\int_0^t d\tau e^{-\frac{x_\tau^2}{4\sigma^2}-\frac{(y_\tau-y_\tau^S)^2}{4{\sigma}^2}}e^{
-i\tau(\nomvk0+\omega_{S,k_{2y}-k_{0y}}-\nomvk2)}\nonumber
\eea
where we have replaced a $\hat\sigma$ with $\sigma$ in $L$ and dropped the corresponding $O(\sigma_0^2/\sigma^2)$ corrections.

Defining the two-vectors
\beq
\vx_0=(x_0,y_0)\hsp
\vec{v}=\left(\frac{k_{0x}}{\nomvk0},\frac{k_{0y}}{\nomvk0}+\frac{k_{0y}-k_{2y}}{\omega_{S,k_{2y}-k_{0y}}}\right)
\eeq
\begin{figure}[htbp]
\centering
\includegraphics[width = 0.6\textwidth]{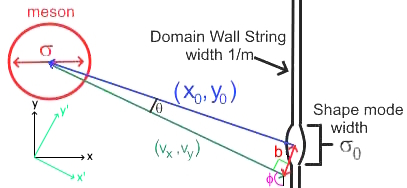}
\caption{The initial conditions for anti-Stokes scattering consist of a meson at position $(x_0,y_0)$ in a round wave packet of size $\sigma$ moving toward a shape mode at the origin, in a wave packet of size $\sigma_0$.  The vector $(v_x,v_y)$ is the initial relative velocity of the meson and the shape mode.  Our treatment allows the initial shape mode to move along the domain wall string, but for simplicity in this picture it is stationary.}\label{fig}
\end{figure}
shown in Fig.~\ref{fig}, we may write
\bea
x_\tau^2+(y_\tau-y_\tau^S)^2&=&\left| \vx_0+\vec{v}\tau\right|^2=|\vec{v}|^2\left(\tau+\frac{\vec{x}_0\cdot \vec{v}}{|\vec{v}|^2}\right)^2+|\vx_0|^2-\frac{\left(\vec{x}_0\cdot \vec{v}\right)^2}{|\vec{v}|^2}\nonumber\\
&=&|\vec{v}|^2\left(\tau+\frac{\vec{x}_0\cdot \vec{v}}{|\vec{v}|^2}\right)^2+|\vx_0|^2\sin^2 \theta \label{x2y2}
\eea
where $\theta$ is the angle between $\vx_0$ and $\vec{v}$.  We conclude that, up to a phase which we will ignore
\beq
L=\frac{2\sigma\sqrt{\pi}}{|\vec{v}|}e^{-\frac{|\vx_0|^2}{4\sigma^2}\sin^2 \theta-\frac{\sigma^2}{|\vec{v}|^2}(\nomvk0+\omega_{S,k_{2y}-k_{0y}}-\nomvk2)^2}. \label{tras}
\eeq

As $|x_0|\gg \sigma$, the support of the first Gaussian is at very small $\theta$ and so we may approximate sin$(\theta)=\theta$ to be the impact parameter $b$ of the collision\footnote{As the wave packet size $\sigma$ is much greater than the width $1/m$ of the domain wall string, $b$ measures a distance perpendicular to $\vec{v}$.  It does not measure a distance parallel to the string.} divided by $|\vx_0|$
\beq
L=\frac{2\sigma\sqrt{\pi}}{|\vec{v}|}e^{-\frac{b^2}{4\sigma^2}-\frac{\sigma^2}{|\vec{v}|^2}(\nomvk0+\omega_{S,k_{2y}-k_{0y}}-\nomvk2)^2}\hsp b=|\vx_0||\sin\theta|.\label{b2}
\eeq
The angle $\theta$ is given by
\beq
\sin \theta = \frac{\vx_0\cdot (-v_y,v_x)}{|\vx_0||\vec{v}|}=\frac{y_0v_x-x_0 v_y}{|\vx_0||\vec{v}|} \label{theq}
\eeq
and so the impact parameter is
\beq
b=\frac{|y_0v_x-x_0 v_y|}{|\vec{v}|}. \label{beq}
\eeq
The impact parameter may be understood as follows.  Let $\tau_c$ be the expected collision time
\beq
x_{\tau_c}=0\hsp \tau_c=-\frac{x_0}{v_x}.
\eeq
If we define the incident angle $\phi$ by
\beq
\tan\phi=\frac{v_x}{v_y}
\eeq
then the impact parameter satisfies
\beq
\frac{b}{\sin\phi}=\left|y_{\tau_C}-y^S_{\tau_C}
\right|=\left| y_0-\frac{x_0 v_y}{v_x}\right|=\left|y_0-\left(\frac{k_{0y}\omega_{S,k_{2y}-k_{0y}}+(k_{0y}-k_{2y})\ok 0}{k_{0x}\omega_{S,k_{2y}-k_{0y}}}\right)x_0\right|.
\eeq

The geometric factor of $\sin(\phi)$ results from the fact that the travel time $\tau$ is longer than $\tau_C$ if $\sin(\phi)\neq 1$.


The final state, up to an overall phase, is then
\bea
e^{-iH\p t}|\Phi\rangle_0\bigg|_{O(\sl)}&=&-\frac{i\sigma\sqrt{\pi}}{2|\vec{v}|}\sqrt{\lambda}\npinvk2 \frac{V_{k_{2x},S,-k_{0x}}}{\nomvk0\omega_{S,k_{2y}-k_{0y}}}\\
&&\ \ \times e^{-\frac{b^2}{4\sigma^2}-\frac{\sigma^2}{|\vec{v}|^2}(\nomvk0+\omega_{S,k_{2y}-k_{0y}}-\nomvk2)^2-\frac{\sigma_0^2\sigma^2}{\hat{\sigma}^2}(k_{0y}–k_{2y})^2}e^{-it\nomvk2}|\vk_2\rangle_0.\nonumber
\eea
In the support of the first Gaussian, $\vx_0$ and $\vec{v}$ are parallel up to corrections of order $O(1/x_0)$.  The second Gaussian is much broader, with a width of $O(1/\sigma)$, and so we may, if desired, insert the parallel condition into the second Gaussian at any point in the calculation.  Any two of the Gaussians together fix both components of $\vk_2/m$ into a range that shrinks to zero in our large $m|x_0|$ and $m\sigma$ limits.

\subsection{Amplitude, Probability and Cross Section}

Our amplitude is 
\beq
\frac{{}_0\langle \vk_2|e^{-iH\p t}|\Phi\rangle_0}{{}_0\langle 0\vac_0}=-\frac{i\sigma\sqrt{\pi\lambda}}{4|\vec{v}|} \frac{V_{k_{2x},S,-k_{0x}}}{\nomvk0\nomvk2\omega_{S,k_{2y}-k_{0y}}} e^{-\frac{b^2}{4\sigma^2}-\frac{\sigma^2}{|\vec{v}|^2}(\nomvk0+\omega_{S,k_{2y}-k_{0y}}-\nomvk2)^2-\frac{\sigma_0^2\sigma^2}{\hat{\sigma}^2}(k_{0y}–k_{2y})^2-it\nomvk2}.
\eeq
Using the inner product
\beq
\frac{{}_0\langle \vk_1\vk_S|\vk_2\vk_S\p\rangle_0}{{}_0\langle 0\vac_0}=\frac{(2\pi)^3\delta(\vk_1-\vk_2)\delta(k_{Sy}-k_{Sy}\p)}{4\nomvk1\omega_{\vk_S}}
\eeq
we obtain the correction factor resulting from the norm of the initial state
\beq
\begin{aligned}
 \frac{{}_0\langle\Phi|\Phi\rangle_0}{{}_0\langle 0\vac_0}
=&\int\frac{dk_{Sy}\p}{2\pi}e^{-\sigma_0^{2}k_{Sy}^{'2}}\int\frac{dk_{Sy}}{2\pi}e^{-\sigma_0^{2}k_{Sy}^2}\npinvk1\npinvk2\alpha_{\vk_1}\alpha^*_{\vk_2}\frac{{}_0\langle Sk_{Sy}\vk_2|Sk_{Sy}^{'}\vk_1\rangle_0}{{}_0\langle 0\vac_0}\\
=&\int\frac{dk_{Sy}}{2\pi}e^{-2\sigma_0^{2}k_{Sy}^2}\pinvk\frac{|\alpha_{\vk}|^2}{4\omega_{\vk_S}\omvk}
=\int\frac{dk_{Sy}}{2\pi}\frac{e^{-2\sigma_0^{2}k_{Sy}^2}}{4\omega_{\vk_S}\omega_{\vk_0}}\pinvk{|\alpha_{\vk}|^2}\\
=&\frac{1}{4\omega_{S}\omega_{\vk_0}}\frac{1}{2\sqrt{2\pi}\sigma_0}
\npinvk1
(2\sigma\sqrt{\pi})^4e^{-2\sigma^2(k_{0x}-k_{1x})^2}
e^{-2\sigma^2(k_{0y}-k_{1y})^2}\\
=&\frac{\sqrt{2}\pi^{3/2}\sigma^4}{\sigma_0\omega_{S}\omega_{\vk_0}}\left(\frac{1}{2\sqrt{2\pi}\sigma} 
\right)^2=\frac{\sqrt{\pi}\sigma^2}{4\sqrt{2}\sigma_0\omega_{S}\omega_{\vk_0}}.
\end{aligned}
\eeq


We want to calculate the probability that the final state has one ground state domain wall and one meson.  Such states are preserved by the projector
\beq
\mathcal{P}=\int d^2\vk_2 \mathcal{P}_{\rm{diff}}(\vk_2)\hsp  \mathcal{P}_{\rm{diff}}(\vk_2)=\frac{1}{(2\pi)^2} \frac{2\nomvk2|\vk_2\rangle_0{}_0\langle \vk_2|}{{}_0\langle 0\vac_0} .
\eeq

Finally we may assemble all of these ingredients to write the total probability $P_{\rm{aS}}(\vk_0)$
of Stokes scattering at $O(\lambda)$ 
\beq
P_{\rm{aS}}=\int d^2\vec{k}_2 P_{\rm{aS}}(\vk_2)_{\rm{diff}}
=\frac{{}_0\langle\Phi|e^{iH\p t}\mathcal{P}e^{-iH\p t}|\Phi\rangle_0}{{}_0\langle\Phi|\Phi\rangle_0}
=\npinvk2 \frac{2\nomvk2}{{}_0\langle 0\vac_0}\frac{\left|{}_0\langle\vk_2|e^{-iH\p t}|\Phi\rangle_0\right|^2}{{}_0\langle\Phi|\Phi\rangle_0}
\eeq
where the differential probability is
\bea
P_{\rm{aS}}(\vk_2)_{\rm{diff}}&=&
\frac{2\nomvk2}{{}_0\langle 0\vac_0}\frac{\left|{}_0\langle\vk_2|e^{-iH\p t}|\Phi\rangle_0\right|^2}{{}_0\langle\Phi|\Phi\rangle_0}\\
&=&{\sqrt{\frac{\pi}{2}}\sigma_0\omega_{S}}{}\frac{{\lambda}}{|\vec{v}|^2} \frac{|V_{k_{2x},S,-k_{0x}}|^2}{\nomvk0\nomvk2\omega_{S,k_{2y}-k_{0y}}^2} e^{-\frac{b^2}{2\sigma^2}-\frac{2\sigma^2}{|\vec{v}|^2}(\nomvk0+\omega_{S,k_{2y}-k_{0y}}-\nomvk2)^2-\frac{2\sigma_0^2\sigma^2}{\hat{\sigma}^2}(k_{0y}–k_{2y})^2}.
\nonumber
\eea
Note that the 3-point interaction $V$ varies slowly with $k_{2x}$, and so in our $\sigma m\rightarrow\infty$ limit, over the narrow $k_2$-support of the Gaussian factors the variation of $V$ tends to zero.  Therefore we may remove the $V$ term from the $\vk_2$ integral, setting $k_2$ to its value at the Gaussian peaks.  Actually, there are two such peaks $\pm k_{2x}^I$ where we have defined
\beq
\omega_{\vk_2^I}=\nomvk0+\omega_S\hsp k_{2x}^I=\sqrt{(\nomvk0+\omega_S)^2-k_{0y}^2-m^2}. \label{k2i}
\eeq
At these two respective peaks, the vertex factor $V$ becomes $V_{\pm k_{2x}^I,S,-k_{0x}}$.

We would now like to perform the $\vk_2$ integrations.  These are constrained by all three Gaussians.  The first Gaussian, constraining the impact parameter, is the narrowest.  This is because only a slight deviation in the incoming velocity can cause the meson to miss, as $|x_0|\gg\sigma$.  Therefore, to understand the dependence of the probability on the impact parameter, we should integrate that Gaussian first.

While this is straightforward and the results may be interesting, we will instead turn our attention to the total cross section.  For this purpose, a shift in $\vk_2$ affects $b$ but that shift can be undone with a shift in $y_0$ and so is inconsequential for the total cross section.  Therefore, we will simply ignore the $\vk_2$ dependence in $b$, and pull that Gaussian out of the $\vk_2$ integral.  The total probability is then
\bea
P_{\rm{aS}}&=&
{\sqrt{\frac{\pi}{2}}\sigma_0}{}\frac{{\lambda}}{|\vec{v}|^2} \frac{\left(|V_{k^I_{2x},S,-k_{0x}}|^2+|V_{-k^I_{2x},S,-k_{0x}}|^2\right)}{\nomvk0\nomvk2^I\omega_{S}}\nonumber\\
&&\times e^{-\frac{b^2}{2\sigma^2}}\npinvk2 e^{-\frac{2\sigma^2}{|\vec{v}|^2}(\nomvk0+\omega_{S,k_{2y}-k_{0y}}-\nomvk2)^2-\frac{2\sigma_0^2\sigma^2}{\hat{\sigma}^2}(k_{0y}–k_{2y})^2}. \label{totp}
\eea
Here we have included a factor of two for the two peaks, and so it is understood that the integral is to be performed over just one peak, using the Gaussian approximation which we will now describe.

To perform the Gaussian integrations, notice that although the first Gaussian is more narrow, the coefficient nonetheless will be evaluated near the peak of the second where $|\vec{v}|^2=|\vk_0|^2/\nomvk0^2$.  

Now the two-dimensional Gaussian integral may be performed by taking the determinant of the Hessian matrix.

In the case of the $\sigma_0$ term, the eigenvalue is $-4\sigma_0^2\sigma^2/\hat{\sigma}^2$ and the eigenvector is $k_{2y}$.  The determinant is the product of this eigenvalue with the second $k_{2x}$ derivative of the other term
\beq
-\frac{2\sigma^2\nomvk0^2}{|\vk_0|^2}\frac{\partial^2}{\partial_{k_{2x}}^2}(\nomvk0+\omega_{S,k_{2y}-k_{0y}}-\nomvk2)^2=-\frac{4\sigma^2\nomvk0^2}{|\vk_0|^2}\frac{k^{I2}_{2x}}{\omega_{\vk_2^I}^{2}}.
\eeq
Note that there are actually two degenerate peaks, corresponding to $k_{2x}=\pm k^I_{2x}$, leading to an additional factor of 2.

In all, at each of the two peaks, we find
\beq
\npinvk2 e^{-\frac{2\sigma^2}{|\vec{v}|^2}(\nomvk0+\omega_{S,k_{2y}-k_{0y}}-\nomvk2)^2-\frac{2\sigma_0^2\sigma^2}{\hat{\sigma}^2}(k_{0y}–k_{2y})^2}=\frac{1}{4\pi} \frac{\hat{\sigma}}{\sqrt{2}\sigma_0\sigma}\frac{
|\vk_0|(\nomvk0+\omega_S)}{\sqrt{2}\sigma k^I_{2x}\nomvk0}.
\eeq
Up to corrections of order $O(\sigma^2_0/\sigma^2)$ we may set the $\hat\sigma/\sigma$ factor to unity.  The probability of anti-Stokes scattering is then
\beq
P_{\rm{aS}}=\frac{\lambda}{8\sqrt{2\pi}\sigma}
\frac{|V_{k^I_{2x},S,-k_{0x}}|^2+|V_{-k^I_{2x},S,-k_{0x}}|^2}{|\vk_0| \omega_{S}k^I_{2x}} e^{-\frac{b^2}{2\sigma^2}} .
\eeq
This result cannot be directly compared with that of Ref.~\cite{hengyuanstokes} because here we have considered $\sigma\gg\sigma_0$ and the opposite limit was taken there, however one may observe that if we restrict the present result to the case $k_{0y}=b=0$ and that result to reflectionless potentials, then the answers agree with $\sigma$ and $\sigma_0$ exchanged. Note that, in general, we expect the relative smearing to be $\sqrt{\sigma^2+\sigma_0^2}$, and so one expects that, whenever one of these widths is much larger, the $\sqrt{\sigma^2+\sigma_0^2}$ will be equal to the larger width.

\subsection{Cross Section}

To obtain the cross section $\Sigma_{\rm{aS}}$ for anti-Stokes scattering, we define the coordinates $x\p$ and $y\p$ which are respectively parallel and perpendicular to the $\vec{v}$ direction
\beq
x\p=x\sin\phi+y\cos\phi\hsp
y\p=-x\cos\phi+y\sin\phi.
\eeq
Now the cross section standard formula (4.60) from Peskin and Schroeder \cite{ps}, in our case is
\beq
P_{\rm{aS}}=\Sigma_{\rm{aS}}l_m l_S \int dy\p \rho_m(y\p)\rho_S(y\p)    \label{peskin}
\eeq
where $l_m$ and $l_S$ are the lengths of the meson and shape mode wave packets in the $x\p$ direction and $\rho_m$ and $\rho_S$ are the respective densities per unit $y\p$ of the meson and shape mode.  We know that there is a single shape mode and a single meson, and so
\beq
l_m\int dy\p \rho_m(y\p)=l_S\int dy\p \rho_S(y\p)=1.
\eeq
The limit $\sigma/\sigma_0\gg 1$ implies that the meson is much wider than the shape mode, and so we may approximate
\beq
l_S\rho_S(y\p)=\delta(y\p-b)
\eeq
to be a Dirac $\delta$ function and so we obtain
\beq
P_{\rm{aS}}=\Sigma_{\rm{aS}} l_m\rho_m(b).\label{rhom}
\eeq
Matching the $b$-dependence of both sides, and recalling that $l_S\rho_S$ integrates to unity, one finds, unsurprisingly, that
\beq
l_m\rho_m(b)=\frac{e^{-\frac{b^2}{2\sigma^2}}}{\sqrt{2\pi}\sigma}.\label{rhom1}
\eeq
Intuitively this density may be understood as follows.  We have ignored the smearing of the final wave packet caused by the spread in the momenta.  We have only considered the smearing of the original position $|\vx_0|$.  Holding $\vec{v}$ fixed, our formula (\ref{beq}) tells us that the smearing in $b$ is that of the initial meson wave packet along the direction perpendicular to $\vec{v}$.  Our initial meson wave packet enjoyed a rotational symmetry, with a spread of $\sigma$ on each axis, and so the spread in $b$ is also $\sigma$.

We then conclude that the cross section for anti-Stokes scattering is
\beq
\Sigma_{\rm{aS}}=\frac{\lambda}{8}
\frac{\left(|V_{k^I_{2x},S,-k_{0x}}|^2+|V_{-k^I_{2x},S,-k_{0x}}|^2\right)}{|\vk_0| \omega_{S}k^I_{2x}}. \label{princ}
\eeq
This is our main result, where we remind the reader that $k_{2x}^I$ was reported in Eq.~(\ref{k2i}).  Note that the cross section depends on $\vk_0$.  Unlike the probability, it is independent of $x_0$, $y_0$, $\sigma$ and $\sigma_0$.  When $k_{0y}=0$ this cross section is equal to the probability of anti-Stokes scattering on a kink in 1+1 dimensions, as calculated in Ref.~\cite{mestokes}.  The cross section in 2+1 dimensions has dimensions of length and the probability is always dimensionless, consistent with the fact that in d+1 dimensions $\lambda$ has dimensions of [mass${}^{3-d}$].

We observe that, as $\rho_m$ has a simple Gaussian form of width $\sigma$, one may have arrived at the cross section more simply.  One could have computed the probability at $b=0$ as in Ref.~\cite{hengyuanstokes}, but with $\sigma\gg\sigma_0$, and then multiplied by $\sqrt{2\pi}\sigma$. 

\subsection{A Broad Shape Mode}

What about the other limit, $\sigma_0\gg\sigma$, in which the shape mode wave packet is broader than the meson wave packet?  Let us return to Eq.~(\ref{noas}), before we imposed that the meson wave packet was broader.

We may simplify the phases in $L$ somewhat by using the first order expansions
\beq
\omega_{k_{0x}k_{2y}}=\omega_{\vk_0}+(k_{2y}-k_{0y})\frac{k_{0y}}{\omega_{\vk_0}}\hsp
\omega_{S0}=\omega_{S,k_{2y}-k_{0y}}-(k_{2y}-k_{0y})\frac{k_{2y}-k_{0y}}{\omega_{S,k_{2y}-k_{0y}}}
\eeq
to rewrite the phases as
\beq
\tau\left(\omega_{\vk_0}+\omega_{S,k_{2y}-k_{0y}}-\omega_{\vk_2}\right)+(k_{2y}-k_{0y})(y_\tau-y_\tau^S)=(k_{2y}-k_{0y})y_0+\tau\left(\omega_{k_{0x}k_{2y}}+\omega_{S0}-\omega_{\vk_2}\right).
\eeq

\begin{figure}[htbp]
\centering
\includegraphics[width = 0.6\textwidth]{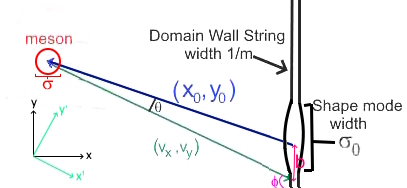}
\caption{As in Fig.~\ref{fig} except now, $\sigma_0\gg\sigma$.  As a result, the collision occurs when the meson is roughly coincident with the domain wall.  This means that the impact parameter measures a distance along the wall, and not perpendicular to the velocity.}\label{fig2}
\end{figure}

Rewriting $L$ as
\beq
L=\int_0^t d\tau e^{M}
\eeq
then we may complete the square of the dominant terms, those with $\sigma^2$ in the denominator, in the exponent
\bea
M&=&-\frac{v_x^2}{4\sigma^2}\left[
\tau+\frac{x_0}{v_x}+i\frac{2\sigma^2}{v_x^2}\left(\omega_{k_{0x}k_{2y}}+\omega_{S0}-\omega_{\vk_2}\right)\right]^2+i\frac{x_0}{v_x}\left(\omega_{k_{0x}k_{2y}}+\omega_{S0}-\omega_{\vk_2}\right)\nonumber\\
&&-\frac{\sigma^2}{v_x^2}\left(\omega_{k_{0x}k_{2y}}+\omega_{S0}-\omega_{\vk_2}\right)^2-\frac{v_y^2}{4\sigma_0^2}\tau^2-\frac{v_yy_0}{2\sigma_0^2}\tau-\frac{y_0^2}{4\sigma_0^2}-i(k_{2y}-k_{0y})y_0. \label{meq}
\eea
The term in the square brackets is a narrow Gaussian.  As the other terms vary slowly over the support of this Gaussian, we simply evaluate them at the peak
\beq
\overline{\tau}=-\frac{x_0}{v_x}-i\frac{2\sigma^2}{v_x^2}\left(\omega_{k_{0x}k_{2y}}+\omega_{S0}-\omega_{\vk_2}\right)
\eeq
of the Gaussian and pull them out of the integral where their real parts are
\beq
{\rm{Re}}\left[-\frac{v_y^2}{4\sigma_0^2}\overline{\tau}^2-\frac{v_yy_0}{2\sigma_0^2}\overline{\tau}-\frac{y_0^2}{4\sigma_0^2}\right]=-\frac{1}{4\sigma_0^2}\left(y_0-\frac{v_yx_0}{v_x}\right)^2+\frac{\sigma^4v_y^2}{\sigma_0^2v_x^4}\left(\omega_{k_{0x}k_{2y}}+\omega_{S0}-\omega_{\vk_2}\right)^2.
\eeq
Now $v_x^2\sigma_0/v_y\gg \sigma$, because we take the limit $\sigma/\sigma_0\rightarrow 0$ with $v_x$ and $v_y$ fixed, so the last term is negligible compared with the first term in the last line of Eq.~(\ref{meq}).
We then find, up to a phase that we will ignore
\beq
L=\frac{2\sigma\sqrt{\pi}}{v_x}\exp{-\frac{b^2}{4\sigma_0^2}-\frac{\sigma^2}{v_x^2}\left(\omega_{k_{0x}k_{2y}}+\omega_{S0}-\omega_{\vk_2}\right)^2}
\eeq
where we have defined the impact factor
\beq
b=\left|y_0-\frac{v_y}{v_x}x_0\right|.
\eeq
Note that, unlike the case $\sigma\gg \sigma_0$, in the present case the interaction occurs when the meson strikes the wall, and so the impact factor measures a distance along the $y$ axis and not orthogonal to $\vec{v}$.  This situation is shown in Fig.~\ref{fig2}.

Note that this result for $L$ is quite similar to the other case (\ref{b2}) except that $|\vec{v}|$ has been replaced by $v_x$, the frequencies are evaluated at different $y$ momenta and the definition of $b$ has been changed.  Also, the factor of $\hat\sigma$ in the denominator of Eq.~(\ref{noas}) now should be approximated to $\sigma_0$ instead of $\sigma$.  Recall that this factor is squared in arriving at the probability, and so one gains a factor of $\sigma^2/\sigma_0^2$.  As a result, the calculation of the probability (\ref{totp}) proceeds as before, now leading to
\bea
P_{\rm{aS}}&=&
\sqrt{\frac{\pi}{2}}\frac{\sigma^2}{\sigma_0}\frac{{\lambda}}{v_x^2} \frac{\left(|V_{k^I_{2x},S,-k_{0x}}|^2+|V_{-k^I_{2x},S,-k_{0x}}|^2\right)}{\nomvk0\nomvk2^I\omega_{S}}\nonumber\\
&&\times e^{-\frac{b^2}{2\sigma_0^2}}\npinvk2 e^{-\frac{2\sigma^2}{v_x^2}\left(\omega_{k_{0x}k_{2y}}+\omega_{S0}-\omega_{\vk_2}\right)^2-\sigma^2(k_{0y}–k_{2y})^2}. 
\eea
Integrating over $\vk_2$ as before, the two Hessian eigenvalues are now $-4\sigma^2$ and
\beq
-\frac{4\sigma^2\omega_{\vk_0}^2}{k_{0x}^2}\frac{k_{2x}^{I2}}{\omega_{\vk^I_2}^2}.
\eeq
Performing a Gaussian integration on $\vk_2$ and summing over the sign of $k^I$ then yields
\beq
P_{\rm{aS}}=\frac{\lambda}{8\sqrt{2\pi}\sigma_0}
\frac{|V_{k^I_{2x},S,-k_{0x}}|^2+|V_{-k^I_{2x},S,-k_{0x}}|^2}{k_{0x} \omega_{S}k^I_{2x}} e^{-\frac{b^2}{2\sigma^2_0}} .
\eeq
The calculation of the cross section is a bit different from the case in which the meson was more smeared.  For one, now we approximate the meson density to be a $\delta$ function.   But there is a more important distinction.  Let $(x_0\p,y_0\p)$ be the initial position $(x_0,y_0)$ in the $(x\p,y\p)$ coordinates.  Then
\beq
b=\left|y_0-\cot(\phi)x_0\right|=\left|\frac{y\p_0}{\sin(\phi)}\right|.
\eeq
The $y_0\p$ dependence in $P_{aS}$ is then
\beq
e^{-\frac{b^2}{2\sigma_0^2}}=e^{-\frac{y^{\prime 2}}{2\sigma_0^2\sin(\phi)}}.
\eeq
Again this needs to be matched to the density $l_S\rho_S(y\p)$, which is normalized to unity, leading to
\beq
l_S\rho_S(b)=\frac{e^{-\frac{b^2}{2\sigma_0^2}}}{\sqrt{2\pi}\sigma_0\sin(\phi)}.
\eeq
Intuitively, the additional factor of $\sin(\phi)$ in the denominator came from the fact that the target shape mode is essentially a line, and so the interaction time from oblique scattering includes a determinant factor which depends on the incident angle.

This leads to an additional factor of $\sin(\phi)$ in the cross section.  Using the identity
\beq
k_{0x}=|\vec{k}|\sin(\phi)
\eeq
we obtain the same cross section (\ref{princ}) as in the case $\sigma\gg\sigma_0$.

\section{Example: $\phi^4$ Double-Well Model} \label{fsez}

\begin{figure}[htbp]
\centering
\includegraphics[width = 0.6\textwidth]{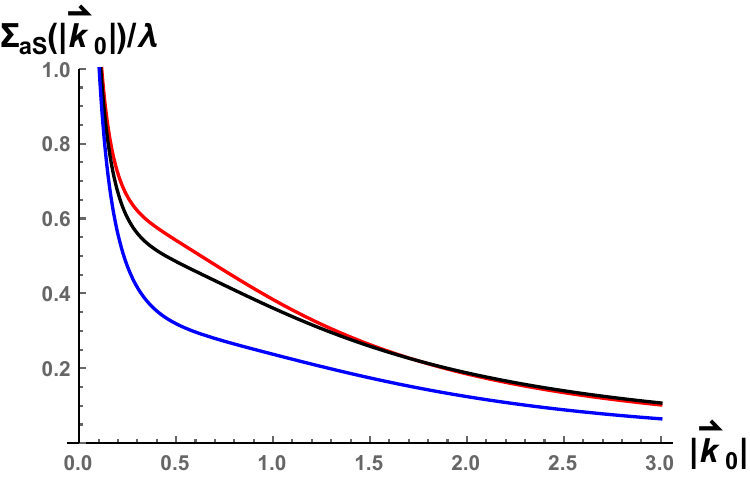}
\caption{The cross section for anti-Stokes scattering as a function of the momentum $\vk_0$ of the incident meson, with units chosen such that $m=1$.  The incident angle $\phi$ is $90^\circ$ (red), $60^\circ$ (black) and $30^\circ$ (blue).}\label{sfig}
\end{figure}

In 2+1 dimensions, only a $\phi^4$ or $\phi^6$ potential will lead to a renormalizable theory.  Both admit domain walls, although in general sextic potentials have vacua in which the mesons have different masses and so the walls accelerate.  For simplicity, we will consider the $\phi^4$ double-well model which is characterized by the potential
\beq
V(\sqrt{\lambda}\phi(\vx))=\frac{\lambda\phi^2(\vx)}{4}\left(\sqrt{\lambda}\phi(\vx)-\sqrt{2}m\right)^2.
\eeq
The domain wall enjoys a single shape mode
\beq
g_S(x)=\frac{\sqrt{3m}}{2}\tanh\left(\frac{mx}{2}\right)\sech\left(\frac{mx}{2}\right)\hsp \omega_S=\frac{\sqrt{3}m}{2}.
\eeq
The three-point coupling is
\beq
V_{Sk_1k_2}=\pi\frac{3\sqrt{3}}{32\sqrt{2}}\frac{\left(17m^4-16(k_1^2-k_2^2)^2\right)(m^2+4k_1^2+4k_2^2)+128m^2k_1^2k_2^2}{m^{3/2}\ok{1}\ok{2}\sqrt{m^2+4k_1^2}\sqrt{m^2+4k_2^2}}\sech\left(\frac{\pi(k_1+k_2)}{m}\right).\label{v12}
\eeq
Substituting this into our master formula (\ref{princ}) for the cross section, we obtain an analytic formula for the cross section for any initial momentum $\vk_0$.  In Fig.~\ref{sfig} this cross section is plotted for various incident angles.

\section{Remarks}
In the derivation of the probability of anti-Stokes scattering, we have made two unjustified approximations.  First, the frequency was only expanded to first order in Eq.~(\ref{omex}).  This means that we ignore the smearing of the wave packet as it propagates, which may well dominate the initial smearing in our $|x_0|\gg\sigma$ limit.  The result is that our formula (\ref{theq}) for $\theta$ will be incorrect, and so our formula for $b$ will be underestimated.  The impact parameter will instead have a large variance, with some mesons that began heading toward the shape mode not interacting while others that started not moving toward the shape mode will interact.

Second, in Eq.~(\ref{totp}) we pulled the impact parameter out of the $\vk_2$ integration.  We therefore ignored a second source of smearing, due to the velocity spread.  Again, a proper treatment would have led to a larger variance for the impact parameter $b$.

Both of these poor approximations lead to a prediction for the probability which is too sharply localized as a function of the initial conditions.  However, as in the case of vacuum sector scattering, the localization appears in both sides of Eq.~(\ref{peskin}) and so does not affect the cross section.  Thus we again substantiate the motivation for the present work, that the cross section and not the probability is a more useful quantity in the case of anti-Stokes scattering in 2+1 dimensions, in contrast with the case of Stokes scattering in 2+1 dimensions or any scattering in 1+1 dimensions.

\appendix

\section* {Acknowledgement}

\noindent
This work was supported by the Higher Education and Science Committee of the Republic of Armenia (Research Project Nos. 24PostDoc/2‐1C009 and 24RL-1C047).
HY Guo was supported by Sun Yat-sen university international Postdoctoral Exchange Program and also supported by Research Projects Developed by the Lanzhou Theoretical Physics Center/Gansu Provincial Key Laboratory of Theoretical Physics (Research Project: NSFC Grant No. 12247101).

\end{document}

\bibitem{Kibble:1976sj}
T.~W.~B.~Kibble,
``Topology of Cosmic Domains and Strings,''
J. Phys. A \textbf{9} (1976), 1387-1398
doi:10.1088/0305-4470/9/8/029

\bibitem{bog67}
N. N. Bogoliubov, ``Field-Theoretical Methods in Physics,'' Suppl. al Nuovo Cimento, Vol. IV, serie I (1966) 346-368.

\bibitem{Zeldovich:1974uw}
Y.~B.~Zeldovich, I.~Y.~Kobzarev and L.~B.~Okun,
``Cosmological Consequences of the Spontaneous Breakdown of Discrete Symmetry,''
Zh. Eksp. Teor. Fiz. \textbf{67} (1974), 3-11
SLAC-TRANS-0165.

\bibitem{Dvali:1996xe}
G.~R.~Dvali and M.~A.~Shifman,
``Domain walls in strongly coupled theories,''
Phys. Lett. B \textbf{396} (1997), 64-69
[erratum: Phys. Lett. B \textbf{407} (1997), 452]
doi:10.1016/S0370-2693(97)00131-7
[arXiv:hep-th/9612128 [hep-th]].

\bibitem{Steer:2006ik}
D.~A.~Steer and T.~Vachaspati,
``Domain walls and fermion scattering in grand unified models,''
Phys. Rev. D \textbf{73} (2006), 105021
doi:10.1103/PhysRevD.73.105021
[arXiv:hep-th/0602130 [hep-th]].

\bibitem{muk24}
M.~Mukhopadhyay, O.~Pujolas and G.~Zahariade,
``Cosmological scaling of precursor domain walls,''
Phys. Rev. D \textbf{110} (2024) no.6, 063548
doi:10.1103/PhysRevD.110.063548
[arXiv:2406.10330 [hep-th]].

\bibitem{dan24}
I.~Dankovsky, S.~Ramazanov, E.~Babichev, D.~Gorbunov and A.~Vikman,
``Numerical analysis of melting domain walls and their gravitational waves,''
[arXiv:2410.21971 [hep-ph]].

\bibitem{big24}
F.~Bigazzi, A.~L.~Cotrone and A.~Olzi,
``Cosmic Topological Defects from Holography,''
[arXiv:2411.19302 [hep-th]].

\bibitem{yuki24}
Y.~Amari and M.~Nitta,
``Skyrmion crystal phase on a magnetic domain wall in chiral magnets,''
[arXiv:2409.07943 [cond-mat.mes-hall]].

 \bibitem{yxq}
Yu.~Xiaoquan, Blakie, P. B.
``Dark-soliton-like magnetic domain walls in a two-dimensional ferromagnetic superfluid,''
Phys. Rev. Res.\textbf{3} (2021),023-043
doi:10.1103/PhysRevResearch.3.023043

\bibitem{marra24}
P.~Marra and A.~Nigro,
``Topological zero modes and bounded modes at smooth domain walls: Exact solutions and dualities,''
[arXiv:2408.16466 [hep-th]].

\bibitem{rout24}
A.~Rout and B.~Altschul,
``Bound States and Particle Production by Breather-Type Background Field Configurations,''
[arXiv:2410.09273 [hep-th]].

\bibitem{shiff24}
S.~Chen, E.~Ievlev and M.~Shifman,
``Two types of domain walls in $\mathcal{N}=1$ super-QCD: how they are classified and counted,''
[arXiv:2411.16845 [hep-th]].





\bibitem{Manton:2004tk}
N.~S.~Manton and P.~Sutcliffe,
``Topological solitons,''
Cambridge University Press, 2004,
ISBN 978-0-521-04096-9, 978-0-521-83836-8, 978-0-511-20783-9
doi:10.1017/CBO9780511617034

\bibitem{Lou:2022ksp}
S.~Y.~Lou, X.~z.~Hao and M.~Jia,
``Deformation conjecture: deforming lower dimensional integrable systems to higher dimensional ones by using conservation laws,''
JHEP \textbf{03} (2023), 018
doi:10.1007/JHEP03(2023)018
[arXiv:2211.06844 [nlin.SI]].

\bibitem{kklan}
Y.~Zhong, X.~L.~Du, Z.~C.~Jiang, Y.~X.~Liu and Y.~Q.~Wang,
``Collision of two kinks with inner structure,''
JHEP \textbf{02} (2020), 153
doi:10.1007/JHEP02(2020)153
[arXiv:1906.02920 [hep-th]].

\bibitem{kk0}
H.~Yan, Y.~Zhong, Y.~X.~Liu and K.~i.~Maeda,
``Kink-antiKink collision in a Lorentz-violating $\phi^4$ model,''
Phys. Lett. B \textbf{807} (2020), 135542
doi:10.1016/j.physletb.2020.135542
[arXiv:2004.13329 [hep-th]].

\bibitem{kk1}
M.~Mohammadi and E.~Momeni,
``Scattering of Kinks in the B\ensuremath{\varphi}4 model,''
Chaos Solitons and Fractals: the interdisciplinary journal of Nonlinear Science and Nonequilibrium and Complex Phenomena \textbf{165} (2022), 112834
doi:10.1016/j.chaos.2022.112834
[arXiv:2207.00655 [nlin.CD]].

\bibitem{kk2}
I.~Takyi, S.~Gyampoh, B.~Barnes, J.~Ackora-Prah and G.~A.~Okyere,
``Kink Collision in the Noncanonical $\varphi^{6}$ Model: A Model with Localized Inner Structures,''
[arXiv:2209.05902 [hep-th]].

\bibitem{sfal21}
C.~Adam, D.~Ciurla, K.~Oles, T.~Romanczukiewicz and A.~Wereszczynski,
``Sphalerons and resonance phenomenon in kink-antikink collisions,''
Phys. Rev. D \textbf{104} (2021) no.10, 105022
doi:10.1103/PhysRevD.104.105022
[arXiv:2109.01834 [hep-th]].

\bibitem{car24}
O.~N.~Karp\'\i{}\v{s}ek, L.~Rafaj and F.~Blaschke,
``Scattering of kinks in coreless potentials,''
[arXiv:2407.14313 [hep-th]].

\bibitem{doreyf6}
P.~Dorey, K.~Mersh, T.~Romanczukiewicz and Y.~Shnir,
``Kink-antiKink collisions in the $\phi^6$ model,''
Phys. Rev. Lett. \textbf{107} (2011), 091602
doi:10.1103/PhysRevLett.107.091602
[arXiv:1101.5951 [hep-th]].

\bibitem{mech22}
F.~Blaschke and O.~N.~Karp\'\i{}\v{s}ek,
``Mechanization of scalar field theory in 1+1 dimensions,''
PTEP \textbf{2022} (2022) no.10, 103A01
doi:10.1093/ptep/ptac104
[arXiv:2202.05675 [hep-th]].

\bibitem{yunguo24}
L.~Long, X.~Li and Y.~Jiang,
``A toy model to explain the missing bounce windows in the kink-antikink collisions,''
Phys. Lett. B \textbf{854} (2024), 138763
doi:10.1016/j.physletb.2024.138763
[arXiv:2409.04008 [hep-th]].

\bibitem{f622}
C.~Adam, P.~Dorey, A.~Garcia Martin-Caro, M.~Huidobro, K.~Oles, T.~Romanczukiewicz, Y.~Shnir and A.~Wereszczynski,
``Multikink scattering in the $\phi^6$ model revisited,''
[arXiv:2209.08849 [hep-th]].

\bibitem{nav23}
S.~Navarro-Obreg\'on, L.~M.~Nieto and J.~M.~Queiruga,
``Inclusion of radiation in the collective coordinate method approach of the \ensuremath{\phi}4 model,''
Phys. Rev. E \textbf{108} (2023) no.4, 044216
doi:10.1103/PhysRevE.108.044216
[arXiv:2305.00497 [hep-th]].

\bibitem{sw19}
C.~Adam, K.~Oles, T.~Romanczukiewicz and A.~Wereszczynski,
``Spectral Walls in Soliton Collisions,''
Phys. Rev. Lett. \textbf{122} (2019) no.24, 241601
doi:10.1103/PhysRevLett.122.241601
[arXiv:1903.12100 [hep-th]].

\bibitem{fermwall}
J.~G.~F.~Campos, A.~Mohammadi, J.~M.~Queiruga, A.~Wereszczynski and W.~J.~Zakrzewski,
``Fermionic spectral walls in kink collisions,''
JHEP \textbf{01} (2023), 071
doi:10.1007/JHEP01(2023)071
[arXiv:2211.07754 [hep-th]].

\bibitem{mechris}
J.~Evslin, C.~Halcrow, T.~Romanczukiewicz and A.~Wereszczynski,
``Spectral walls at one loop,''
Phys. Rev. D \textbf{105} (2022) no.12, 125002
doi:10.1103/PhysRevD.105.125002
[arXiv:2202.08249 [hep-th]].

\bibitem{hertzberg}
M.~P.~Hertzberg,
``Quantum Radiation of Oscillons,''
Phys. Rev. D \textbf{82} (2010), 045022
doi:10.1103/PhysRevD.82.045022
[arXiv:1003.3459 [hep-th]].

\bibitem{vachosc}
J.~Oll\'e, O.~Pujolas, T.~Vachaspati and G.~Zahariade,
``Quantum Evaporation of Classical Breathers,''
Phys. Rev. D \textbf{100} (2019) no.4, 045011
doi:10.1103/PhysRevD.100.045011
[arXiv:1904.12962 [hep-th]].

\bibitem{noiosc}
J.~Evslin, T.~Roma\'nczukiewicz and A.~Wereszczy\'nski,
``Quantum oscillons may be long-lived,''
JHEP \textbf{08} (2023), 182
doi:10.1007/JHEP08(2023)182
[arXiv:2305.18056 [hep-th]].

\bibitem{Dashen:1974ci}
R.~F.~Dashen, B.~Hasslacher and A.~Neveu,
``Nonperturbative Methods and Extended Hadron Models in Field Theory 1. Semiclassical Functional Methods,''
Phys. Rev. D \textbf{10} (1974), 4114
doi:10.1103/PhysRevD.10.4114

\bibitem{gjscc}
J.~L.~Gervais, A.~Jevicki and B.~Sakita,
``Collective Coordinate Method for Quantization of Extended Systems,''
Phys. Rept. \textbf{23} (1976), 281-293
doi:10.1016/0370-1573(76)90049-1

\bibitem{Christ:1975wt}
N.~H.~Christ and T.~D.~Lee,
``Quantum Expansion of Soliton Solutions,''
Phys. Rev. D \textbf{12} (1975), 1606
doi:10.1103/PhysRevD.12.1606

\bibitem{vachcqc}
T.~Vachaspati and G.~Zahariade,
``Classical-Quantum Correspondence for Fields,''
JCAP \textbf{09} (2019), 015
doi:10.1088/1475-7516/2019/09/015
[arXiv:1807.10282 [hep-th]].

\bibitem{specrev}
N.~Graham and H.~Weigel,
``Quantum corrections to soliton energies,''
Int. J. Mod. Phys. A \textbf{37} (2022) no.19, 2241004
doi:10.1142/S0217751X22410044
[arXiv:2201.12131 [hep-th]].

\bibitem{gj76}
J.~L.~Gervais and A.~Jevicki,
``Point Canonical Transformations in Path Integral,''
Nucl. Phys. B \textbf{110} (1976), 93-112
doi:10.1016/0550-3213(76)90422-3

\bibitem{rebhan}
A.~Rebhan and P.~van Nieuwenhuizen,
``No saturation of the quantum Bogomolnyi bound by two-dimensional supersymmetric solitons,''
Nucl. Phys. B \textbf{508} (1997), 449-467
doi:10.1016/S0550-3213(97)00625-1
[arXiv:hep-th/9707163 [hep-th]].

\bibitem{Evslin:2023ypw}
J.~Evslin and H.~Liu,
``(Anti-)Stokes scattering on kinks,''
JHEP \textbf{03} (2023), 095
doi:10.1007/JHEP03(2023)095
[arXiv:2301.04099 [hep-th]].

\bibitem{Evslin:2022fzf}
J.~Evslin, H.~Liu and B.~Zhang,
``Meson production from kink-meson scattering,''
Phys. Rev. D \textbf{107} (2023) no.2, 025012
doi:10.1103/PhysRevD.107.025012
[arXiv:2211.01794 [hep-th]].

\bibitem{medomain}
K.~Ogundipe, J.~Evslin, B.~Zhang and H.~Guo,
``A (2+1)-dimensional domain wall at one-loop,''
JHEP \textbf{05} (2024), 098
doi:10.1007/JHEP05(2024)098
[arXiv:2403.14062 [hep-th]].

\bibitem{Jaimungal}
S.~Jaimungal, G.~W.~Semenoff and K.~Zarembo,
``Universality in effective strings,''
JETP Lett. \textbf{69} (1999), 509-515
doi:10.1134/1.568059
[arXiv:hep-ph/9811238 [hep-ph]].

\bibitem{wstabile}
H.~Weigel,
``Quantum Instabilities of Solitons,''
AIP Conf. Proc. \textbf{2116} (2019) no.1, 170002
doi:10.1063/1.5114153
[arXiv:1907.10942 [hep-th]].

\bibitem{cahill76}
K.~E.~Cahill, A.~Comtet and R.~J.~Glauber,
``Mass Formulas for Static Solitons,''
Phys. Lett. B \textbf{64} (1976), 283-285
doi:10.1016/0370-2693(76)90202-1

\bibitem{weigel24}
D.~A.~Petersen and H.~Weigel,
``Vacuum Polarization Energy of a Proca Soliton,''
[arXiv:2411.18373 [hep-th]].

\bibitem{colemanth}
S.~R.~Coleman,
``There are no Goldstone bosons in two-dimensions,''
Commun. Math. Phys. \textbf{31} (1973), 259-264
doi:10.1007/BF01646487

\bibitem{noi31}
J.~Evslin, H.~Liu, B.~Zhang and H.~Guo,
``A Finite Tension for the $\phi^4_4$ Domain Wall,''
[arXiv:2411.05406 [hep-th]].

\bibitem{menorm}
J.~Evslin and H.~Liu,
``A reduced inner product for kink states,''
JHEP \textbf{03} (2023), 070
doi:10.1007/JHEP03(2023)070
[arXiv:2212.10344 [hep-th]].

\end{thebibliography}


\begin{thebibliography}{99}

\bibitem{skyrme}
T.~H.~R.~Skyrme,
``A Nonlinear theory of strong interactions,''
Proc. Roy. Soc. Lond. A \textbf{247} (1958), 260-278
doi:10.1098/rspa.1958.0183

\bibitem{fkk75}
V.~E.~Korepin, P.~P.~Kulish and L.~D.~Faddeev,
``Soliton Quantization,''
JETP Lett. \textbf{21} (1975), 138-139

\bibitem{fk76}
L.~D.~Faddeev and V.~E.~Korepin,
``About the Zero Mode Problem in the Quantization of Solitons,''
Phys. Lett. B \textbf{63} (1976), 435-438
doi:10.1016/0370-2693(76)90390-7

\bibitem{mat77}
V.~A.~Matveev,
``Cancellation of the Zero Mode Singularities in Soliton Quantization Theory,''
Nucl. Phys. B \textbf{121} (1977), 403-412
doi:10.1016/0550-3213(77)90162-6

\bibitem{sol1}
R.~F.~Dashen, B.~Hasslacher and A.~Neveu,
``Nonperturbative Methods and Extended Hadron Models in Field Theory 2. Two-Dimensional Models and Extended Hadrons,''
Phys. Rev. D \textbf{10} (1974), 4130-4138
doi:10.1103/PhysRevD.10.4130

\bibitem{sol2}
J.~Goldstone and R.~Jackiw,
``Quantization of Nonlinear Waves,''
Phys. Rev. D \textbf{11} (1975), 1486-1498
doi:10.1103/PhysRevD.11.1486

\bibitem{sol3}
K.~E.~Cahill, A.~Comtet and R.~J.~Glauber,
``Mass Formulas for Static Solitons,''
Phys. Lett. B \textbf{64} (1976), 283-285
doi:10.1016/0370-2693(76)90202-1


\bibitem{sol4}
L.~D.~Faddeev and V.~E.~Korepin,
``Quantum Theory of Solitons: Preliminary Version,''
Phys. Rept. \textbf{42} (1978), 1-87
doi:10.1016/0370-1573(78)90058-3

\bibitem{cq1}
T.~Vachaspati and G.~Zahariade,
``Classical-quantum correspondence and backreaction,''
Phys. Rev. D \textbf{98} (2018) no.6, 065002
doi:10.1103/PhysRevD.98.065002
[arXiv:1806.05196 [hep-th]].

\bibitem{cq2}
T.~Vachaspati and G.~Zahariade,
``Classical-Quantum Correspondence for Fields,''
JCAP \textbf{09} (2019), 015
doi:10.1088/1475-7516/2019/09/015
[arXiv:1807.10282 [hep-th]].

\bibitem{gw22}
N.~Graham and H.~Weigel,
``Quantum corrections to soliton energies,''
Int. J. Mod. Phys. A \textbf{37} (2022) no.19, 2241004
doi:10.1142/S0217751X22410044
[arXiv:2201.12131 [hep-th]].


\bibitem{gs74}
J.~L.~Gervais and B.~Sakita,
``Extended Particles in Quantum Field Theories,''
Phys. Rev. D \textbf{11} (1975), 2943
doi:10.1103/PhysRevD.11.2943

\bibitem{cl75}
N.~H.~Christ and T.~D.~Lee,
``Quantum Expansion of Soliton Solutions,''
Phys. Rev. D \textbf{12} (1975), 1606
doi:10.1103/PhysRevD.12.1606

\bibitem{tom75}
E.~Tomboulis,
``Canonical Quantization of Nonlinear Waves,''
Phys. Rev. D \textbf{12} (1975), 1678
doi:10.1103/PhysRevD.12.1678

\bibitem{gj76}
J.~L.~Gervais and A.~Jevicki,
``Point Canonical Transformations in Path Integral,''
Nucl. Phys. B \textbf{110} (1976), 93-112
doi:10.1016/0550-3213(76)90422-3

\bibitem{bh24}
A.~Bhattacharya, J.~Cotler, A.~Dersy and M.~D.~Schwartz,
``Collective coordinate fix in the path integral,''
Phys. Rev. D \textbf{110} (2024) no.11, 116023
doi:10.1103/PhysRevD.110.116023
[arXiv:2402.18633 [hep-th]].

\bibitem{ga24}
B.~Garbrecht and N.~Wagner,
``False vacuum decay of excited states in finite-time instanton calculus,''
[arXiv:2412.20431 [hep-th]].

\bibitem{hayashi1}
A.~Hayashi, S.~Saito and M.~Uehara,
``Pion - nucleon scattering in the Skyrme model and the P wave Born amplitudes,''
Phys. Rev. D \textbf{43} (1991), 1520-1531
doi:10.1103/PhysRevD.43.1520

\bibitem{hayashi2}
A.~Hayashi, S.~Saito and M.~Uehara,
``Pion - nucleon scattering in the soliton model,''
Prog. Theor. Phys. Suppl. \textbf{109} (1992), 45-72
doi:10.1143/PTPS.109.45

\bibitem{mekink}
J.~Evslin,
``Manifestly Finite Derivation of the Quantum Kink Mass,''
JHEP \textbf{11} (2019), 161
doi:10.1007/JHEP11(2019)161
[arXiv:1908.06710 [hep-th]].

\bibitem{me2loop}
J.~Evslin and H.~Guo,
``Two-Loop Scalar Kinks,''
Phys. Rev. D \textbf{103} (2021) no.12, 125011
doi:10.1103/PhysRevD.103.125011
[arXiv:2012.04912 [hep-th]].

\bibitem{memult}
H.~Liu, J.~Evslin and B.~Zhang,
``Meson production from domain wall-meson scattering,''
Phys. Rev. D \textbf{107} (2023) no.2, 025012
doi:10.1103/PhysRevD.107.025012
[arXiv:2211.01794 [hep-th]].

\bibitem{mestokes}
J.~Evslin and H.~Liu,
``(Anti-)Stokes scattering on kinks,''
JHEP \textbf{03} (2023), 095
doi:10.1007/JHEP03(2023)095
[arXiv:2301.04099 [hep-th]].

\bibitem{hengyuanstokes}
H.~Guo, H.~Liu and J.~Evslin,
``(Anti-)Stokes scattering on the domain wall string,''
JHEP \textbf{02} (2025), 039
doi:10.1007/JHEP02(2025)039
[arXiv:2412.13409 [hep-th]].

\bibitem{stokes52}
G. G. Stokes, 
``On the Change of Refrangibility of Light,'' Philosophical Transactions of the Royal Society of London 142 (1852) 463–562 doi/10.1098/rstl.1852.0022 

\bibitem{raman28}
C. V. Raman, K. S. Krishnan, ``The Negative Absorption of Radiation,'' Nature 122 (1928), 12–13 doi:10.1038/122012b0

\bibitem{landsberg28} 
G. S. Landsberg, L. I. Mandelstam, ``New Phenomenon in the Scattering of Light (preliminary report),'' Jour. of the Russ. Phys.-Chem. Soc., Phys. 60 (1928) 335.

\bibitem{1703.06696}
M.~Hindmarsh, J.~Lizarraga, J.~Urrestilla, D.~Daverio and M.~Kunz,
``Scaling from gauge and scalar radiation in Abelian Higgs string networks,''
Phys. Rev. D \textbf{96} (2017) no.2, 023525
doi:10.1103/PhysRevD.96.023525
[arXiv:1703.06696 [astro-ph.CO]].

\bibitem{0812.1929}
M.~Hindmarsh, S.~Stuckey and N.~Bevis,
``Abelian Higgs Cosmic Strings: Small Scale Structure and Loops,''
Phys. Rev. D \textbf{79} (2009), 123504
doi:10.1103/PhysRevD.79.123504
[arXiv:0812.1929 [hep-th]].

\bibitem{2405.06030}
A.~Alonso-Izquierdo, J.~J.~Blanco-Pillado, D.~Migu\'elez-Caballero, S.~Navarro-Obreg\'on and J.~Queiruga,
``Excited Abelian-Higgs vortices: Decay rate and radiation emission,''
Phys. Rev. D \textbf{110} (2024) no.6, 065009
doi:10.1103/PhysRevD.110.065009
[arXiv:2405.06030 [hep-th]].



\bibitem{2209.12945}
J.~J.~Blanco-Pillado, D.~Jim\'enez-Aguilar, J.~M.~Queiruga and J.~Urrestilla,
``The dynamics of domain wall strings,''
JCAP \textbf{05} (2023), 011
doi:10.1088/1475-7516/2023/05/011
[arXiv:2209.12945 [hep-th]].

\bibitem{2411.13521}
J.~J.~Blanco-Pillado, A.~Garc\'\i{}a Mart\'\i{}n-Caro, D.~Jim\'enez-Aguilar and J.~M.~Queiruga,
``Effective Actions for Domain Wall Dynamics,''
[arXiv:2411.13521 [hep-th]].

\bibitem{ai25}
W.~Y.~Ai, M.~Carosi, B.~Garbrecht, C.~Tamarit and M.~Vanvlasselaer,
``Bubble wall dynamics from nonequilibrium quantum field theory,''
[arXiv:2504.13725 [hep-ph]].

\bibitem{vinc72}
P.~Vinciarelli,
``Effective mass and correlation length of nucleon constituents,''
Lett. Nuovo Cim. \textbf{4S2} (1972), 905-909
doi:10.1007/BF02756261

\bibitem{cornwall74}
J.~M.~Cornwall, R.~Jackiw and E.~Tomboulis,
``Effective Action for Composite Operators,''
Phys. Rev. D \textbf{10} (1974), 2428-2445
doi:10.1103/PhysRevD.10.2428

\bibitem{erice}
S.~R.~Coleman,
``Classical Lumps and their Quantum Descendents,''
Subnucl. Ser. \textbf{13} (1977), 297
PRINT-77-0088 (HARVARD).

\bibitem{cocorr23}
L.~Berezhiani, G.~Cintia and M.~Zantedeschi,
``Perturbative Construction of Coherent States,''
[arXiv:2311.18650 [hep-th]].

\bibitem{wstabile}
H.~Weigel,
``Quantum Instabilities of Solitons,''
AIP Conf. Proc. \textbf{2116} (2019) no.1, 170002
doi:10.1063/1.5114153
[arXiv:1907.10942 [hep-th]].


\bibitem{wentzel}
G. Wentzel, 
``Zur Paartheorie der Kernkr\"afte,''
Helv. Phys. Acta 15 (1942) 111.


\bibitem{cahill76}
K.~E.~Cahill, A.~Comtet and R.~J.~Glauber,
``Mass Formulas for Static Solitons,''
Phys. Lett. B \textbf{64} (1976), 283-285

\bibitem{noi21}
K.~Ogundipe, J.~Evslin, B.~Zhang and H.~Guo,
``A (2+1)-dimensional domain wall at one-loop,''
JHEP \textbf{05} (2024), 098
doi:10.1007/JHEP05(2024)098
[arXiv:2403.14062 [hep-th]].


\bibitem{ps}
M. E. Peskin and D. V. Schroeder
``An Introduction to quantum field theory,''
Addison-Wesley (1995).






\end{thebibliography}
\end{document}